\documentclass[aps,12pt]{revtex4-1}
\usepackage{graphicx}
\usepackage{amsmath}
\usepackage{esint}
\pdfoutput=1
\usepackage[colorlinks,hyperfigures,unicode]{hyperref}

\begin{document}
\title{Interactions of  Phonons and Rotons with Interfaces\\ in Superfluid Helium}

\author{I.V.~Tanatarov}
\email{igor.tanatarov@gmail.com}
\affiliation{National Science Center ``Kharkov Institute of Physics and Technology'',\\
         Academicheskaya St. 1, Kharkov, 61108, Ukraine}

\author{I.N.~Adamenko}

\author{K.E.~Nemchenko}
\affiliation{Karazin Kharkov National University, Svobody Sq. 4, Kharkov, 61077, Ukraine}

\author{A.F.G.~Wyatt}
\affiliation{School of Physics, University of Exeter, Exeter EX4 4QL, UK}


\date{\today}

\begin{abstract} 
We solve the problem of beams of phonons and rotons incident on, and interacting with, solid surfaces.
Phonons and rotons are the quasiparticles of superfluid helium and have a unique dispersion curve.
The dispersion curve controls the transmission, reflection and mode change of these quasiparticles at the interface with another medium. We develop a non-local hydrodynamic theory in a consistent and unified way. The structure of the solutions in the quantum fluid  is discussed. The creation probabilities of all quasiparticles are derived when any one of them is incident on the interface. The dependencies on frequency and angular are analysed and the backward reflection and refraction for $R^-$ rotons are discussed. 
\end{abstract}

\maketitle 

\section{Introduction}
Many physical properties of continuous media at low temperatures can be described in terms of quasiparticles. The dispersion relation of superfluid helium is non-monotonic and the quasiparticles that correspond to different monotonic regions of the dispersion curve are called phonons, $R^-$ rotons, and $R^+$ rotons (see Fig.\ref{Fig1}). The $R^-$ rotons have negative group velocity, i.e. their momentum is directed opposite to their group velocity. The phonons and rotons are observed in many experiments, such as in neutron scattering in helium \cite{neutron} and in the direct experiments \cite{exp1,exp2}, where beams of superfluid helium quasiparticles are created by a heated solid. The quasiparticles propagate in the helium, interact and reflect from different surfaces and quantum evaporate helium atoms from the free surface. 

These processes have been investigated both experimentally and theoretically (see for example \cite{phononpulses1,phononpulses2,phononpulses3,phononpulses3a,phononpulses4}). Interestingly, $R^-$ rotons were not detected in direct experiments until 1999, when they were finally created by a specially constructed source \cite{exp2} and observed by quantum evaporation. All the earlier attempts to create $R^-$, with ordinary solid heaters and bolometers, were unsuccessful.

He~II quasiparticles interact  with the interface and can cause their transmission, reflection, and conversion into each other (i.e. mode change). These are the fundamental elementary processes that determine heat exchange between He~II and a solid, and cause phenomena such as the Kapitza  temperature jump at a solid-liquid helium interface. (see, for example, Ref. \cite{Khalatnikov}). The phonons are long wavelength fluctuations of density, with the same nature as longitudinal phonons in a solid, and in the long wavelength limit, the problem of their interaction with an interface with a solid, is reduced to the problem of transmission of waves with some constant frequency through this interface. This problem can be found in textbooks on acoustics. The theory can be generalized to take into account the structure of real solids \cite{Khalatnikov} and the structure of the real interfaces (see \cite{AdamenkoSurf1,AdamenkoSurf2,Syrkin}).

For short-wave excitations, rotons, there are still various models and views regarding their nature (see for example \cite{Rot-Feynman}, \cite{Rot-Kruglov} and \cite{Rot-Balibar}), and the problem of their interaction with the interface is of significant interest. The first work in this direction was \cite{Bowley}. However, the method used there did not take into account the possibility of simultaneous creation of phonons and $R^+$ and $R^-$ rotons on the interface, and it could not distinguish between the $R^+$ and $R^-$ rotons.

In the current work we consider the problem of the interaction of superfluid helium quasiparticles with interfaces in a consistent and unified way. We use the approach that describes a quantum fluid as a continuous medium at the length scales that include the average interatomic distance, in terms of dispersive hydrodynamics (see \cite{Pitaevskii,P&S92}). This was developed by the authors in \cite{PRB}. Such an approach is feasible because in a quantum fluid the atoms are delocalised, as the thermal de Broglie wavelength is much larger than $n^{-1/3}$ where $n$ is the atom number density in the liquid.

The idea to describe superfluid helium as a continuous medium at microscopic scales has been successfully used for decades. Atkins \cite{Atkins} used it in the 1950s to describe the mobility of electrons and ions in He~II. In Ref. \cite{Pitaevskii}, nonlocal hydrodynamics was introduced to describe small oscillations in superfluid helium, and in \cite{P&S92} it was used to describe ripplon-roton hybridization. The density functional approach (see for example \cite{DFT1}) also utilizes this idea. However, the theoretical justification of this approach remained on the intuitive level until the work \cite{PRB}. 

The nonlocality in the theory allows one to describe a medium with an arbitrary nonlinear dispersion relation $\Omega(k)$, which appears explicitly in its equations and is the only input "parameter" in the theory. The possibility of using nonlocal hydrodynamics to describe a medium with an arbitrary dispersion relation is discussed for example in \cite{Whitham}. Both phonons and rotons are considered as purely longitudinal excitations of the same nature, only with different wavelengths. The quasiparticles are described as wave packets in a continuous medium  which has nonlocal bonds and nonlinear dispersion. The problem of their reflection or transmission through the interface is reduced to the problem of the interaction of the eigenmodes of the equations of the continuous medium  with the interface.

As an intermediate step to solving the problem for all the quasiparticles of superfluid helium, we considered different special cases of $\Omega(k)$. First, the monotonic dispersion relation $\Omega^{2}\!\sim\! k^{2}\!+\!\alpha k^{4}$, with $\alpha\!>\!0$, was analyzed in one-dimensional \cite{JLTPold,PhNT} and three-dimensional \cite{JLTP2006} cases. In these references, the solutions of the nonlocal equations of dispersive hydrodynamics in the half-space were obtained, and different effects of nonlinearity of $\Omega(k)$ on the problem of waves transmission through the interface were distinguished.

Then in \cite{PRB2008} the dispersion relation such that $\Omega^{2}(k^{2})$ is a cubic polynomial of $k^{2}$ was considered. It is the simplest expression that can approximate the distinctive dispersion relation of superfluid helium, in both the phonon and roton regions. The problem of the simultaneous creation of phonons and rotons on the interface was solved and the interaction of all He II quasiparticles with the interface was described in a unified way. The failures of attempts to detect $R^-$ rotons, before the experiments of Ref. \cite{exp2}, were explained, and predictions were made for new experiments on the interaction of phonons and rotons with a solid. However, the use of a simple cubic approximation for the dispersion relation of superfluid helium restricted the results obtained in \cite{PRB2008} to qualitative or semi-quantitative.

In the current work we present the consistent solution of the problem of the interaction of quasiparticles with an interface for the case when their dispersion relation is arbitrary and non-monotonic, so that $\Omega^{2}(k^{2})$ is a polynomial of arbitrarily large degree $S$. The probability of the creation of each quasiparticle at the interface is derived for all cases. This work includes and generalizes the results of works \cite{JLTPold}-\cite{PRB2008}, and in the  special cases discussed previously, the expressions obtained here transform into the ones obtained earlier. So all results are now presented in a unified way. 

The exact expressions for the quasiparticles' creation probabilities on the interface are derived analytically. Those expressions are valid for all energies and angles of incidence of the quasiparticles that are incident on the interface. They are also valid for monotonic and non-monotonic dispersion relation, and for interfaces with a solid or for the free surface. All the qualitative results of \cite{PRB2008} with regard to superfluid helium, including the explanation why $R^-$ rotons were not detected in direct experiments until \cite{exp2}, are confirmed.

The dispersion relation of surface excitations of He~II, ripplons, can also be investigated in the same framework, but is not considered in this work, and is the subject of another investigation \cite{Ripplon}.

In the next section we consider the equations of the quantum fluid in a half-space and the boundary conditions that apply on the interface. Section 3 is devoted to the solution of the equations, and in section 4 we discuss some general consequences of the solution's structure and of the use of boundary conditions with regard to the problem of waves interacting with interfaces. Those include a generalization of Snell's law and a realization of backward reflection and refraction for $R^-$ rotons.

In section 5 we consider the process of a phonon in the solid which is incident on the interface, and derive the reflection and transmission coefficients. In the next section we calculate the partial transmission coefficients for the waves created in quantum fluid (for the case when there are more than one), which are the creation probabilities of the corresponding quasiparticles created in the process. In section 7 we derive all the creation probabilities in the process of a quasiparticle of the quantum fluid incident on the interface, i.e. phonon or $R^-$ roton or $R^+$ roton for the dispersion of superfluid helium. We discuss the peculiarities of the coefficients' frequency and angular dependencies and refine the curves, obtained in \cite{PRB2008} for the coefficients calculated in a rough approximation.

\section{Equations and boundary conditions for He~II\\ in a half-space}

Let us consider the half-space $z\!>\!0$ filled by a quantum fluid with equilibrium density $\rho_{q}$ and dispersion relation $\Omega(k)$. According to the approach of \cite{PRB}, we introduce the variables of continuous medium: the velocity $\mathbf{v}$, and the deviations of pressure $P$ and density $\rho$ from their respective equilibrium values. These quantities must obey the laws of mass and momentum conservation; therefore the excitations of small amplitude are described by ordinary linearised equations of an ideal liquid,  but the relation between $\rho$ and $P$ is non-local, defined by some difference kernel $h(\mathbf{r})$. The problem in terms of pressure $P$ can be expressed as a nonlocal wave equation. When solving the problem in the half-space, the integration domain is limited to this half-space \cite{PRB2008} and the problem can be brought to the form 
\begin{equation}\label{EQP}
	\triangle P(\mathbf{r},t)=
		\!\int\limits_{z_{1}>0}\! d^{3}r_1\,
		 h(|\mathbf{r}\!-\!\mathbf{r}_1|)\ddot{P}(\mathbf{r}_1,t), \quad
		x,y,t\!\in\!(0,\infty),\;z\!\in\!(0,\infty). 
\end{equation}
 where the dots denote derivatives with respect to time. We assume that the interface is sharp enough to consider that the kernel $h(r)$ is the same in the presence of the interface as it is in the bulk medium. 

In the infinite medium, the integration and definition domains of Eq. (\ref{EQP}) are infinite, its right-hand part is a convolution, and the kernel is related to the dispersion relation of the bulk excitations $\Omega(k)$ through its Fourier transform (see \cite{PRB})
\begin{equation}
	 h(k)=\frac{k^{2}}{\Omega ^{2}(k)}. 
\end{equation}

Thus the equation (\ref{EQP}) describes a continuous medium with dispersion relation $\Omega(k)$, which fills the half-space $z\!>\!0$. We do not try to find $\Omega(k)$ from microscopics, but rather use the experimentally measured curve as the ``input parameter'' of the theory and thus determine the kernel $h(r)$. This approach implicitly contains the inner structure of the fluid. The dispersion $\Omega(k)$ can be arbitrary. In particular, we are interested in the case, when it approximates to the distinctive dispersion relation of superfluid helium (see Fig.\ref{Fig1}).

The equation (\ref{EQP}) is supplemented by the boundary conditions on the interface $z\!=\!0$. As the local equations of continuous medium hold on both sides of the interface, the boundary conditions, also local, are obtained from their integral forms in the usual way, using the theory of a continuous medium.

If we consider the interface with a solid, the boundary conditions will have the form 
\begin{equation} \label{BC}
\begin{array}{l}
	\left.P\right|_{z=-0}=\left.P\right|_{z=+0},\\
	\left.\mathrm{v}_{z}\right|_{z=-0}=\left.\mathrm{v}_{z}\right|_{z=+0},
\end{array} 
\end{equation} 
where $\mathrm{v}_{z}$ is the $z$-component of the velocity of the continuous medium.

If we consider the free surface of superfluid helium, with surface tension $\sigma$, then the pressure at the surface should be the Laplace pressure, and for small deviations of the surface from equilibrium position can be written as 
\begin{equation} \label{boundary-landau}
	 P=\sigma
	\left(\frac{\partial^{2}\xi}{\partial x^{2}}+
		\frac{\partial^{2}\xi}{\partial y^{2}}\right),
\end{equation} 
where $\xi$ is the $z$-coordinate of the points of the surface (see for example \cite{P&S92} or \cite{L&L}).

The real interface between helium and a solid or helium and vacuum is, of course, not infinitely sharp; for quasiparticles, the interface can only approximately be considered flat since their wave-length may be comparable with the size of the microstructure of the solid surface. This especially concerns rotons, as their wave-length is comparable with the interatomic distances. The possible effects of taking this into account are discussed at the end of section 7.

\section{Solution of the equation for pressure in the half-space}

\begin{figure}[!ht]
\center
\includegraphics[width=8.6cm, viewport=77 255 417 522]{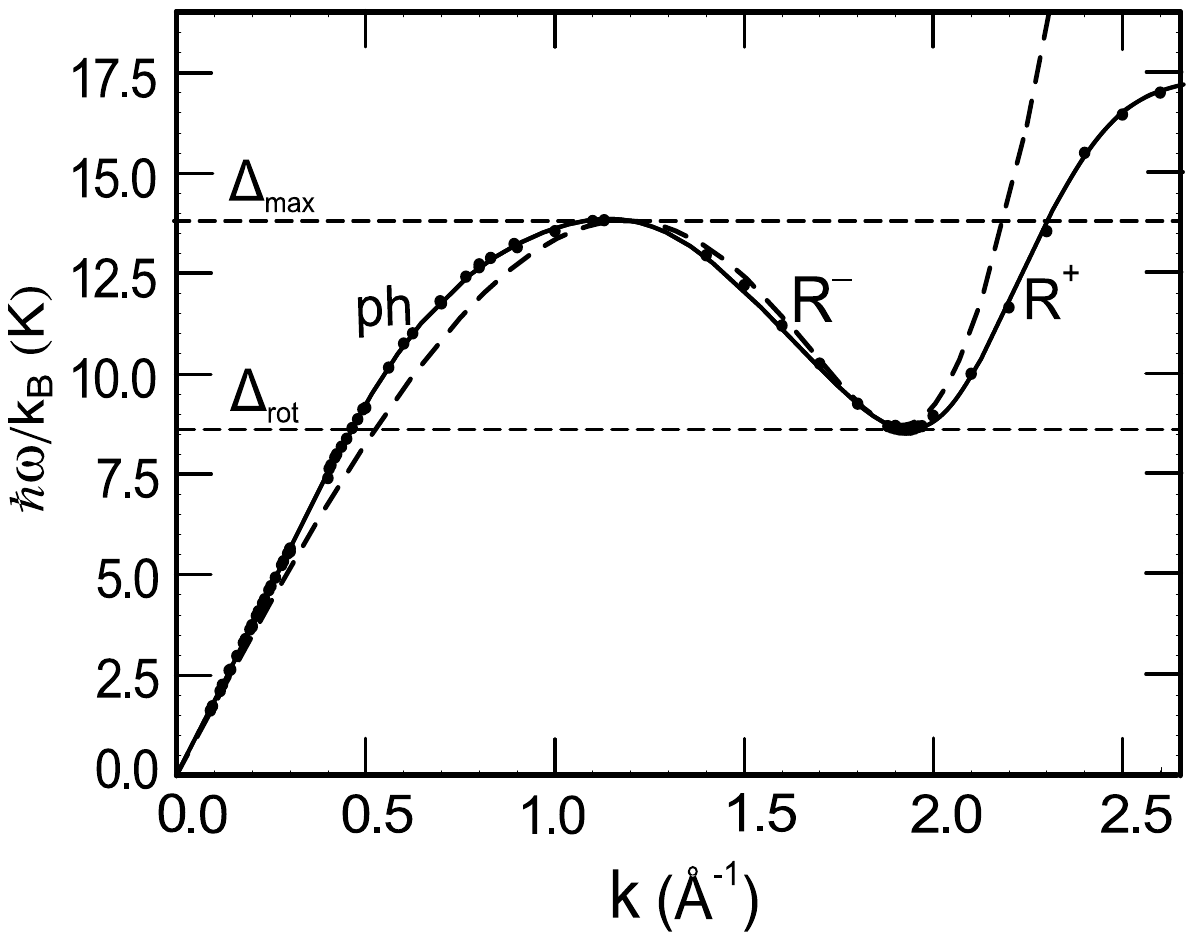}
\begin{minipage}[b]{0.85\textwidth}\caption{\label{Fig1}
Dispersion relation of the quasiparticles of superfluid helium $\Omega(k)$. 
 Dots are experimental data \cite{neutron}, and the solid line
shows $\Omega(k)$ obtained by fitting $\Omega^{2}(k^{2})/k^{2}$ as function of
$k^{2}$ by a polynomial of degree $20$, so $\Omega^{2}(k^{2})$ is polynomial of
degree $S\!=\!21$. The dashed line shows the dispersion curve analysed in
\cite{PRB2008}, with $S\!=\!3$.}
\end{minipage}
\end{figure}

The equation (\ref{EQP}) was solved with the help of the Wiener and Hopf method in
\cite{JLTPold,PhNT} in the one-dimensional case, for the situation when the
function $\Omega^{2}(k^{2})$ is an arbitrary but monotonic polynomial at
$k\!>\!0$; then generalized in \cite{JLTP2006} to three dimensions. In Ref.
\cite{PRB2008} the case of nonmonotonic $\Omega^{2}(k^2)$ of degree $S\!=\!3$
was analyzed in detail. The nonmonotonic cubic polynomial is the simplest
function that can approximate the distinctive dispersion relation of superfluid
helium. Therefore the use of this approximation allowed the authors in
\cite{PRB2008} to solve the problem of superfluid helium phonons and rotons
interaction with an interface, and in particular to derive the reflection and
transmission coefficients for them in terms of roots of the corresponding cubic
equation. However, due to the simplicity of the approximation, the results obtained  were limited to
semi-quantitative, especially for energies close to and higher than maxon
the energy range below energies where the third degree polynomial starts to deviate from the $R^+$
roton part of helium spectrum significantly (see Fig.\ref{Fig1}).

In the current work we use the solution of (\ref{EQP}) for $\Omega^{2}(k^2)$ a
polynomial of arbitrary power. The physical consequences of $\Omega(k)$ being
essentially nonlinear were analyzed in \cite{PhNT,JLTP2006}. Those include the
presence in the solution of exponentially damped waves that correspond to
complex roots of equation $\Omega^{2}(k)=\omega^{2}$ with regard to $k$;
complex  amplitude coefficients of transmission and reflection, the
frequency dependencies of the coefficients, the angles of transmission and
angles of full internal reflection. The consequences of non-monotonicity of
$\Omega(k)$ were studied in detail in \cite{PRB2008} and these include, among
others, multiple critical angles and backward reflection for $R^-$ rotons. Now
we have accumulated enough understanding of the problem to show what happens in
the general case.

The Wiener and Hopf method (see for example \cite{Gahov}, \cite{Mittra&Lee})
can be used to solve the Eq. (\ref{EQP}) for the function $\Omega(k)$ of rather
general form. Its idea is to transform to new functions, to turn the
right-hand part of (\ref{EQP}) into a convolution and after the Fourier
transform by $\mathbf{r}$ and $t$, to reduce the problem to the homogeneous
Riemann boundary value problem (see for example \cite{Riemann}) in the plane of
complex variable $k_{z}$ (see appendix for more detail). The problem is
completely determined by its ``density" 
\begin{equation} \label{densityG}
G(\omega,k)=\frac{\Omega^{2}(k)-\omega^{2}}{\Omega^{2}(k)},
 \end{equation}
where $\omega$ is frequency, $\mathbf{k}=\mathbf{k}_{\tau}+\mathbf{e}_{z}k_{z}$
is the wave vector and $\mathbf{k}_{\tau}$ its projection on the plane $(x,y)$.

The solution of the Riemann problem is standard when $G$ is differentiable and
does not become zero on the real axis of variable $k_z$ \cite{Gahov}. However it can be generalised 
for our case when it has zeros in the real roots of
equation \begin{equation} \label{DispEq}
\Omega^{2}(k^{2}\!=\!k_{\tau}^{2}+k_{z}^{2})=\omega^2 
\end{equation}
with regard to $k_{z}$, and is always zero when $k_{\tau}\!=\!0$ (see
appendix).

The simplest way to bypass the first limitation is to shift the real roots into
the complex plane, while preserving the index of density $G$. The index of $G$
is the key parameter of the Riemann boundary problem, which can be calculated
as the difference between the number of plain zeros of Eq. (\ref{DispEq}) in the
upper half-plane $\mathcal{C}_{+}$ of variable $k_{z}$ and the lower half-plane
$\mathcal{C}_{-}$. In our case $G(k_{z})$ is an even function and the index is
zero. Different ways of shifting the roots lead to different boundary problems
(more details below).

The general solution is rather complicated. In order to make the inverse Fourier transforms and ultimately derive the transmission and reflection coefficients in analytic form, we assume that the function $\Omega^{2}(k^2)$ is a polynomial of power $S$ with regard to $k^{2}$, such that the only real zero of $\Omega^{2}(k^2)$ is $k^{2}\!=\!0$, where $\Omega^{2}\sim k^2$. This assumption actually does not restrict the generality of our consideration, because any given dispersion curve, measured in experiment, can be approximated, on a finite interval of $k$, by a polynomial, with any given precision. The condition that the only real zero of $\Omega^{2}(k^2)$, when $k$ is zero, implies that
there are no other singularities of $G$ on the real line.

For arbitrary degree $S$, it can be shown that the 
 Fourier transform of the solution of equation (\ref{EQP}) in terms of $\mathbf{r}$ and $t$ is
\begin{equation} \label{Solution} 
	 P_{\mu}(k_z;\omega,\mathbf{k}_{\tau})=
	 C_{\mu}(\omega,\mathbf{k}_{\tau})
	\frac{\prod_{\mu}\!\!\!'\left[k_{z}\!-\!k_{i\,z}(0,\mathbf{k}_{\tau})\right]}
		{\prod_{\mu}\left[k_{z}\!-\!k_{i\,z}(\omega,\mathbf{k}_{\tau})\right]}.
\end{equation} 
The product is taken over all the roots
$k_{z}\!=\!k_{i\,z}(\omega,k_{\tau})$ of Eq. (\ref{DispEq}) in the upper
half-plane $\mathcal{C}_{+}$ of the complex variable $k_{z}$. The real roots
are assumed to be shifted from the real line, with the condition that there are
as many roots shifted up as there are shifted down (this comes from the demand
that index of the density is zero). The choice which of the roots are shifted
up (or, equivalently, down) is denoted by the index $\mu$. It determines the
specific Riemann boundary problem, of which (\ref{Solution}) is the solution,
and determines $P_{\mu}$ to within its amplitude
$C_{\mu}(\omega,\mathbf{k}_{\tau})$, because the solution of a Riemann boundary
problem with zero index is unique to within a multiplicative constant
\cite{Gahov}. The linear combination of solutions $P_{\mu}$ with all possible
selections $\mu$, gives the general solution of (\ref{EQP}) with given $\omega$
and $\mathbf{k}_{\tau}$.

As Eq. (\ref{DispEq}) is a polynomial equation of degree $S$ with real
coefficients, either with regard to $k^{2}$ or to $k_{z}^{2}$, the full number
of roots $k_{i\,z}$ in $\mathcal{C}_{+}$, after shifting, is $S$. At
$\omega\!=\!0$, one of them is $\hat{k}_{z}=i|k_{\tau}|$. The prime superscript
on the product in the nominator designates that the factor with this root
$(k_{z}\!-\!\hat{k}_{z})$ is omitted from it. This is the consequence of the
singularity of density $G$ in $k_{z}\!=\!0$ when $k_{\tau}\!=\!0$.

In the limit of small frequencies, when the dispersion relation is almost
linear $\Omega(k)\approx ks$, each factor in the lower product in
(\ref{Solution}), tends to the corresponding factor in the upper product, with
the exception of the factor with $\hat{k}_{z}$, which is absent in
the nominator, so (\ref{Solution}) tends to \begin{equation} \label{SolLinear}
P_{\mu}(k_{z})=\frac{C_{\mu}}{k_{z}-\hat{k}_{z}}. \end{equation}

Suppose that $k_{\tau}\!>\!\omega/s$. Then the two roots of Eq. (\ref{DispEq})
$\pm\sqrt{\omega^{2}/s^{2}-k_{\tau}^{2}}$ are imaginary, and there is no choice
of roots shifting: due to the condition $\hat{k}_{z}\!\in\!\mathcal{C}_{+}$ we have
$\hat{k}_{z}\!=\!i|\sqrt{k_{\tau}^{2}\!-\!\omega^{2}/s^{2}}|$. The solution has
the form (\ref{Solution}) and is unique. It describes the low-frequency surface
excitation of He~II, ripplon $P_{rippl}(z)\sim e^{-|\hat{k}_{z}|z}$.

Now suppose that $k_{\tau}\!<\!\omega/s$. Then of the two roots
$\pm\sqrt{\omega^{2}/s^{2}-k_{\tau}^{2}}$, one should be shifted up. The choice
$\hat{k}_{z}=+|\hat{k}_{z}|$ gives us the solution
$P_{out}(z)\sim\exp{(i|\hat{k}_{z}(\omega,k_{\tau})|z)}$, which is the wave
traveling away from the interface, while the other choice gives the wave
traveling towards the interface
$P_{in}(z)\sim\exp{(-i|\hat{k}_{z}(\omega,k_{\tau})|z)}$. So in this
case there are two linear-independent solutions, and the subscript takes two
values $\mu=out,in$. The two solutions describe low-frequency phonons traveling
in the positive and negative directions of the axis $z$.

In the general case, the solution in  coordinate space, is obtained through
the inverse Fourier transform, by integrating over the upper half-plane
$\mathcal{C}_{+}$ of $k_{z}$. If we are interested in solutions with a given
$\omega$ and $\mathbf{k}_{\tau}$, we take the amplitude of the form
$C_{\mu}(\omega',\mathbf{k}_{\tau}')\!=\!C_{\mu}\delta(\omega-\omega')
\delta(\mathbf{k}_{\tau}-\mathbf{k}_{\tau}')$ and then the solution is the sum of $S$
monochromatic waves 
\begin{equation} \label{SolutionR}
	 P_{\mu}(\mathbf{r},t;\mathbf{k}_{\tau},\omega)=
	 \sum\limits_{k_{iz}\in\mathcal{C}_{+}}
	\alpha_{\mu,i}e^{i(k_{iz}(\omega,k_{\tau})z+\mathbf{k}_{\tau}\mathbf{r}_{\tau}-\omega t)}, 
\end{equation} 
where $\mathbf{r}_{\tau}$ is the projection of the
radius-vector to the plane of the interface $(x,y)$, and the sum is taken over
all the $S$ roots $k_{i\,z}$ in $\mathcal{C}_{+}$ (after shifting). The
amplitudes $\alpha_{\mu,i}$ are the residues of the right-hand part of
(\ref{Solution}) in $k_{i\,z}$.

The velocity is obtained from (\ref{SolutionR}) through the usual relation
$\dot{\mathbf{v}}=-\nabla P/\rho_{q}$: 
\begin{equation} \label{SolutionR-V}
	\mathbf{v}_{\mu}(\mathbf{r},t;\mathbf{k}_{\tau},\omega)=
	\frac{1}{\rho_{q}\omega}\sum\limits_{k_{iz}\in\mathcal{C}_{+}}
	\mathbf{k}_{i}\alpha_{\mu,i}
		 e^{i(k_{i\,z}(\omega,k_{\tau})z+\mathbf{k}_{\tau}\mathbf{r}_{\tau}-\omega t)}. 
\end{equation}
Here $\mathbf{k}_{i}=\mathbf{k}_{\tau}+\mathbf{e}_{z}k_{i\,z}$.

Real solutions, which correspond to quasiparticles propagating in the medium,
are obtained by making wave packets of $P_{\mu}$, i.e. taking the inverse
Fourier transform of Eq. (\ref{Solution}) with the amplitude
$C_{\mu}(\omega',\mathbf{k}_{\tau}')$, that is a function that is essentially
different from zero only when the arguments are in the neighborhood of given
$\omega$ and $\mathbf{k}_{\tau}$. Such solutions consist of $S$ wave packets,
some of them traveling and some of them damped in $z\!>\!0$.

If, for some given $\omega$ and $k_{\tau}$, there are $N$ positive roots
$k_{i\,z}^{2}$ of Eq. (\ref{DispEq}), it means that there are $N$ types of
traveling waves in the quantum fluid (e.g. $R^+$ or $R^-$ rotons). On the other
hand, then $N$ of $2N$ real roots $k_{i\,z}$ will be shifted up from the real
line. Each of the roots $k_{i\,z}$ in the upper half-plane of $k_{z}$ gives a
summand in (\ref{SolutionR}) and (\ref{SolutionR-V}), so the solution $P_{\mu}$
in the quantum fluid will contain $N$ traveling waves and $S\!-\!N$ damped
waves. In terms of the problem of quasiparticles' interaction with the
interface, as will be shown below in more detail, this leads to the following:
when any quasiparticle is incident, all the quasiparticles that can be created
on the interface, which conserve energy and tangential component of momentum,
will be created with corresponding non-zero probabilities.

\section{Transmission of waves through the interface:\\ general properties}

In the next sections we consider the problem of quasiparticles' interaction
with the interface, -- their reflection, transmission and mode change. We treat
quasiparticles as wave packets, so, as the problem is linear, it is reduced to
the problem of plane waves interaction with the interface. However, when
calculating energy transmission coefficients, we have to take into account that
energy (as any other quantity quadratic on amplitude) is carried in a wave
packet with its group velocity.

The solutions both in $z\!>\!0$ and $z\!<\!0$ consist of waves with given
$\omega$ and $\mathbf{k}_{\tau}$. Therefore in order for the boundary
conditions to be satisfied for all $t$ in all the points of the interface
$(x,y)$, the frequencies $\omega$ and tangential components of wave vectors
$\mathbf{k}_{\tau}$ on both sides of the interface must be equal. In terms of
quasiparticles this means that energy and the tangential component of momentum are
conserved in the processes of quasiparticle destruction and creation at the
interface. This leads to several general consequences.

\begin{itemize} 
\item The wave vectors of all the incident, transmitted and
reflected waves lie in one plane, which we may denote by $(y,z)$, and thus the
problem is reduced to two-dimensional. The same consideration applies to the
case of the free surface. 
\item $R^-$ rotons have negative group velocity
$d\Omega/dk\!<\!0$, or "negative dispersion", i.e. their group and phase
velocities are antiparallel. Therefore for them the effects of backward
reflection and refraction are realized, which were predicted and described by
Mandelstam \cite{Mandelshtamm} for a hypothetical (at that time) fluid with
negative group velocity. If the direction of propagation along the plane
$(x,y)$ of all the quasiparticles with positive group velocities (i.e. all
except $R^-$ rotons), incident or created on the interface, is along the vector
$\mathbf{k}_{\tau}$, the $R^-$ rotons will propagate in the opposite direction.
The effect is analogous to Andreev reflection of the quasiparticles on the
interface of normal and superconductive phases of a supercunductor, with change
of signs of electric charge and effective mass \cite{Andreev}. Two examples are
shown on Fig.\ref{Fig2}.

\begin{figure}[htb]
\center
\includegraphics[width=8.6cm, viewport=93 316 513 513]{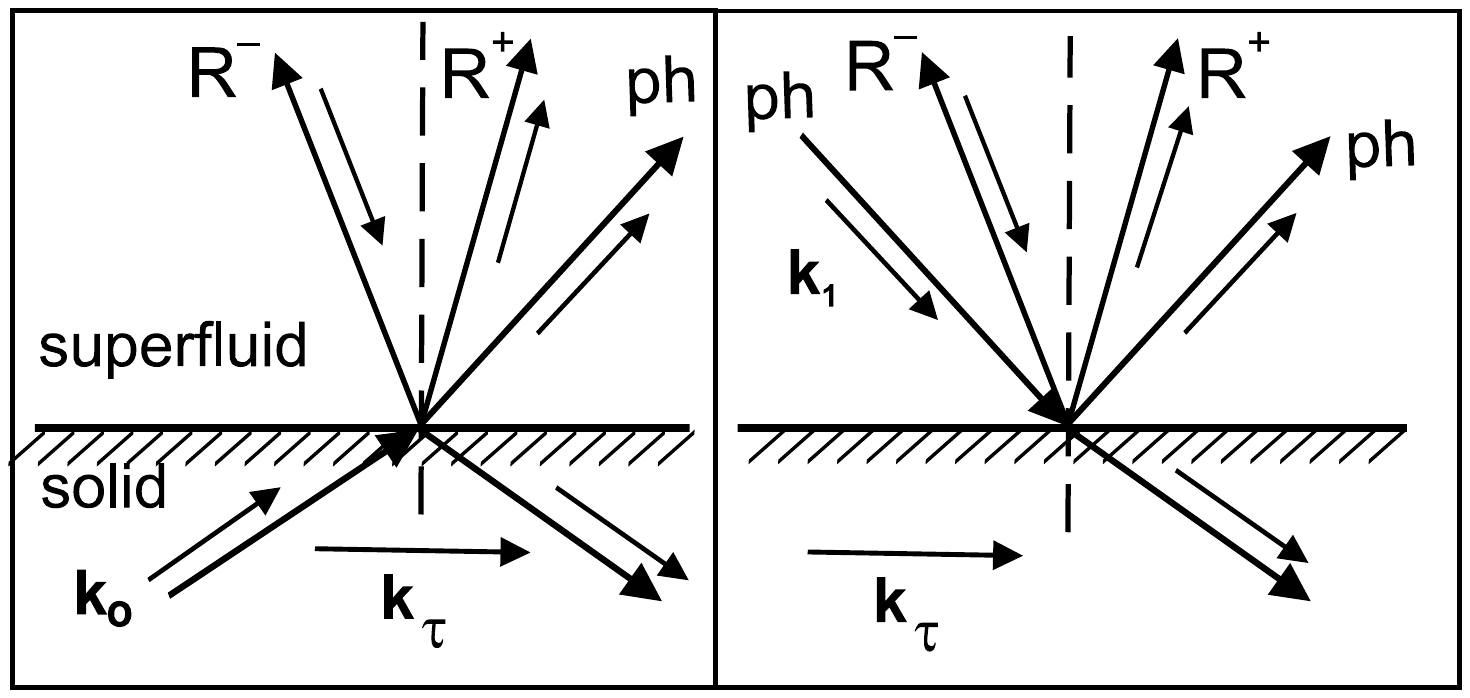}
\begin{minipage}[b]{0.85\textwidth}\caption{\label{Fig2} Two examples of backward refraction (on the left) and reflection (on the right) of $R^-$ rotons. The scheme on the left shows a phonon in the solid incident on the interface, and on the right a phonon of the quantum fluid is incident. The directions of propagation of the quasiparticles (along their group velocities) are shown by big arrows, and the small arrows beside them are their wave vectors. The directions are determined by the condition that the projections of all wave vectors onto the plane of the interface are equal to $\mathbf{k}_{\tau}$.}
\end{minipage}
\end{figure}

\item A form of Snell's law holds for all the traveling waves
involved in the process -- incident, reflected and transmitted, so that the
quantity 
\begin{equation} \label{Snell}
	\frac{\sin\Theta_{i}}{s_{i}(\omega)} 
\end{equation} 
is the same
for all $i$. Here index $i$ enumerates the types of traveling waves with
different phase velocities in the two adjacent media (e.g. helium $R^+$ rotons
or solid's phonons); $s_{i}(\omega)=\omega/k_{i}(\omega)$ is the phase velocity
of wave $i$ and $\Theta_{i}\!>\!0$ is its angle of propagation, measured from
the normal. Thus all the angles of incidence, reflection and transmission are
related through (\ref{Snell}), and it also determines the number of traveling
waves in each process.

Suppose wave $i$ is incident on the interface. Then if for some $j\!\neq\!i$
Eq. (\ref{Snell}) gives $\sin\Theta_{i}\!>\!1$, it means that the corresponding
wave is damped and the quasiparticle $j$ is not created. 
\end{itemize}

Let us assign the subscripts $i=1,2,3$ to the three types of traveling waves in
the superfluid -- the phonons, $R^-$ rotons and $R^+$ rotons. So $k_{1,2,3}$
are the positive roots of Eq. (\ref{DispEq}) with regard to $k$, with the
subscripts in ascending order of their absolute values 
\begin{equation}
	\label{RealRoots}
	0<k_{1}(\omega)<k_{2}(\omega)<k_{3}(\omega). 
\end{equation}

In this work we focus on the peculiarities that the special dispersion relation
of superfluid helium stipulates with regard to the problem of quasiparticles'
interaction with the interface. So, for simplicity, we describe the solid by
a scalar model, i.e. as an isotropic continuous medium with equilibrium density
$\rho_{s}$ and one sound velocity $s_{0}$, and thus assign the index $i=0$ in
Eq. (\ref{Snell}) to the solid's phonons.

As transverse waves can be treated in the same theoretical framework, 
the theory  of  continuous  medium, they do not present any difficulty. 
However,  the  calculations  become much more cumbersome, while,  on  the  whole,
 the  situation does not change. Due to the very small transmission  coefficient of the
solid-helium interface (see section 7), the reflection  coefficients  hardly 
change  at  all. For the transmitted waves  additional  critical  angles 
appear, corresponding to the sound velocity of the transverse waves. Also it
should be noted that taking into  account  both the longitudinal and transverse
waves in the solid, allows  one to consider the contribution of Rayleigh waves,
which give contributions to the transmission coefficients of He~II
quasiparticles into  the  solid  at  fixed  incidence angles. For phonons with
linear dispersion this problem was solved in \cite{Rayleigh}.

\section{Phonon in the solid incident on the interface} 
Let us consider the problem of transmission of a phonon in the solid, with frequency $\omega$, incident from the solid side, on the interface at angle $\Theta_{0}$ to the normal, into superfluid helium.

In the scale of frequencies of superfluid helium's dispersion relation, the dispersion of the solid can be safely regarded as linear and its sound velocity as constant $s_{0}\!=\!const$. In the scalar model the solution in the half-space $z\!<\!0$ is the superposition of the incident and reflected waves with wave vectors $k_{0}\!=\!\omega/s_{0}$. The amplitude reflection coefficient, defined as the ratio of pressures in the reflected and incident waves, can be expressed as 
\begin{equation} \label{r(Z)} 
	 r=\frac{Z-1}{Z+1},
\end{equation} 
where 
\begin{equation} \label{Z}
	 Z=\frac{\cos\Theta_{0}}{\rho_{s}s_{0}}
	\cdot\left.\frac{P}{\mathrm{v}_{z}}\right|_{z=-0} 
\end{equation} 
is a generalization of the wave impedence of the interface. Due to the boundary conditions (\ref{BC}), the ratio $P/\mathrm{v}_{z}$ at $z=-0$ is equal to its value at $z=+0$ and therefore is fully determined by the solution in the quantum fluid, in $z\!>\!0$. This is the quantity, through which all the properties of the quantum fluid affect the reflection coefficient $r$ and the solution in the solid.

Let us now construct the "out-solution" $P_{out}$ in $z\!>\!0$, which is realized when a wave is incident on the interface from the solid. This implies that a wave packet of solutions $P_{out}$ should contain only traveling wave packets that travel away from the interface, if there are any. This construction of the out-solution implies the correct choice $\mu$ of roots $k_{i\,z}$ in $\mathcal{C}_{+}$.

$N$ is the number of positive solutions of Eq. (\ref{DispEq}) with regard to $k_{z}^{2}$ for the given $\omega$ and $k_{\tau}$. Then among the $S$ summands of (\ref{SolutionR}) there are $N$ with real $k_{i\,z}$ and $S\!-\!N$ with $\mathrm{Im}\,k_{i\,z}\!>\!0$. Indeed, of the $2N$ real roots of Eq. (\ref{DispEq}) with regard to $k_{z}$ half would be shifted up from the real line, and thus be included in (\ref{SolutionR}). As $\Omega(k)$ approximates the dispersion relation of superfluid helium, the possible cases to be considered are $N=0,1,2,3$. These cases might, in fact, include all the problems of interest for arbitrary dispersion.

The three roots $k_{1,2,3}$ defined in (\ref{RealRoots}) are all real when $\omega\!\in\!(\Delta_{rot},\Delta_{max})$, where $\Delta_{rot}$ and $\Delta_{max}$ are the roton and maxon frequencies respectively (see Fig.\ref{Fig1}). At $\omega\!<\!\Delta_{rot}$ and $\omega\!>\!\Delta_{max}$ we define them by continuity. Thus $k_{1}$ corresponds to phonons, $k_{2}$ to $R^-$ rotons and $k_{3}$ to $R^+$ rotons in the whole frequency range. $k_{2,3}$ at $\omega\!<\!\Delta_{rot}$ and $k_{1,2}$ at $\omega\!>\!\Delta_{max}$ are complex, among the other $S-1$ complex roots of Eq. (\ref{DispEq}) with regard to $k$ for $\omega$ in those intervals.

Each root $k_{i}$ for $i=1,2,3$ gives a pair $\pm\sqrt{k_{i}^{2}-k_{\tau}^{2}}$ of roots of Eq. (\ref{DispEq}) with regard to $k_{z}$. Of each pair either both roots are real or the two are complex conjugate. Let us denote by $k_{i\,z}$ for $i=1,2,3$ the three roots out of the six that enter the solution $P_{out}$, and show that there is only one way to choose the triplet.

Indeed, of a complex conjugate pair, only the root in $\mathcal{C}_{+}$ can enter (\ref{Solution}), as $P(z)$ should be bounded at $z\!>\!0$: $k_{i\,z}=\sqrt{k_{i}^{2}-k_{\tau}^{2}}\in\mathcal{C}_{+}$. Of the real pair, one of the roots corresponds to a wave traveling towards the interface, and the other, to the wave traveling away from the interface.

$R^-$ rotons have negative group velocity, i.e. for them $d\Omega/dk<0$. Therefore, a wave packet constructed of plane waves with wave vectors close to $k_{2}$ and their $z$-components positive, will travel towards the interface, and vice versa. As the out-solution can only contain wave packets traveling away from the interface, we finally obtain the following rules for choosing $k_{i\,z}$ for $i=1,2,3$: 
\begin{equation} \label{Kz-signs} 
	\begin{array}{l}
	 k_{i\,z}^{2}=k_{i}^{2}(\omega)-k_{\tau}^{2}\quad\mbox{for}\; i=1,2,3;\\
	\mathrm{sign}\, k_{i\,z}=(-1)^{i+1}\quad\mbox{if}\;k_{i}^{2}>k_{\tau}^{2};\\
	k_{i\,z}\in\mathcal{C}_{+}\quad\mbox{if}\;k_{i}^{2}<k_{\tau}^{2}. 
	\end{array}
\end{equation} 
All the other roots of Eq. (\ref{DispEq}) with regard to $k_{z}$ are complex and break up into complex-conjugate pairs. Of each pair there is only one root that lies in $\mathcal{C}_{+}$, and those are the roots $k_{i\,z}$ for $i\!=\!4,\ldots,S$. Then $P_{out}$ solution is built in a definite way and is unique, with $\hat{k}_{z}\!=\!k_{1\,z}$. It has the form (\ref{Solution}) with the real roots shifted up defined by Eq. (\ref{Kz-signs}). The rule includes all the cases $N=0,1,2,3$.

The sound velocities of most metals (e.g. Cu or Au) are much greater than the phase velocity of helium $s(k)=\Omega(k)/k$, so at the interface with superfluid helium it usually holds that $k_{\tau}\!<\!k_{0}=\omega/s_{0}\ll\omega/s(k)\leq k_{1,2,3}$. Therefore the case $N=3$ is realized and we have 
\begin{equation} \label{KzSeq}
	0<k_{1\,z}<(-k_{2\,z})<k_{3\,z} 
\end{equation}

So the out-solution contains three traveling waves, which means that a phonon in the solid, which is incident on the interface, creates in the helium, with non-zero probability, a phonon,  a $R^-$ roton and a $R^+$ roton. In the general case $N$ depends on the function $\Omega(k)$, values of $\omega$, $k_{\tau}$ and $s_{0}$.

In order to obtain the impedance $Z$, we have to calculate the values of $P$ and $\mathrm{v}_{z}$ at $z=0$. From (\ref{SolutionR}), $\left.P\right|_{z=0}$ is the sum of all residues $\alpha_{out,i}$ of the right-hand part of (\ref{Solution}), as the latter all lie in $\mathcal{C}_{+}$, and therefore is equal to minus the residue of $P(k_{z})$ at infinity. From (\ref{SolutionR-V}), $\left.\mathrm{v}_{z}\right|_{z=0}$ is the sum of residues of $k_{z}P(k_{z})/(\rho_{q}\omega)$ in $k_{i\,z}$, and therefore we get
\begin{equation} \label{ResInfty} 
\begin{array}{l}
	\left.P\right|_{z=+0}=-
	\underset{k_{z}\rightarrow\infty}{\mathrm{res}}
	 P_{out}(k_{z});\\
	\left.\mathrm{v}_{z}\right|_{z=+0}=-\frac{1}{\rho_{q}\omega}\cdot
	\underset{k_{z}\rightarrow\infty}{\mathrm{res}}
	\left[k_{z}P_{out}(k_{z})\right].
\end{array}
\end{equation}

The residues at infinity are obtained directly by expanding (\ref{Solution}) and on substituting them into (\ref{ResInfty}) and (\ref{Z}), we obtain
\begin{equation} \label{Zfinal} 
	 Z=\frac{\rho_{q}}{\rho_{s}}\cdot\frac{k_{0\,z}}{k_{1\,z}+
	\sum\limits_{i=2}^{S}\left[k_{i\,z}-k_{i\,z}(\omega\!=\!0)\right]}, 
\end{equation} 
where $k_{0\,z}=k_{0}\cos\Theta_{sol}\!>\!0$ is the $z$-component of the wave vector of the incident wave.

The function of $k_{i\,z}$ in the denominator of Eq. (\ref{Zfinal}) determines the influence of the non-linearity of the dispersion relation $\Omega(k)$ on the reflection and transmission coefficients. In the case of linear dispersion, it is reduced to $k_{1\,z}=\omega/s\,\cdot\cos\Theta_{q}$, where $\Theta_{q}$ is the transmission angle defined by (\ref{Snell}). Then $Z$ turns into the ordinary impedance $\rho_{q}s/\rho_{s}s_{0}$, multiplied by the function of angles $\cos\Theta_{0}/\cos\Theta_{q}$.

If we introduce the notation 
\begin{equation} \label{Knew} 
	\begin{array}{l}
	\tilde{k}_{s}=\frac{\rho_{q}}{\rho_{s}}k_{0\,z};\\
	\tilde{k}_{q}=k_{1\,z}+
	\sum\limits_{i=2}^{S}\left[k_{i\,z}-k_{i\,z}(\omega\!=\!0)\right], 
	\end{array} 
\end{equation} 
then the expression for the amplitude reflection coefficient takes the compact form 
\begin{equation} \label{rFin}
	 r=\frac{\tilde{k}_{s}-\tilde{k}_{q}}{\tilde{k}_{s}+\tilde{k}_{q}}.
\end{equation}

The transmission coefficient is $D=1-|r|^{2}$ and therefore 
\begin{equation}\label{D}
	 D(\omega,k_{\tau})=\frac{4\tilde{k}_{s}\mathrm{Re}\tilde{k}_{q}}
	{\left|\tilde{k}_{s}+\tilde{k}_{q}\right|^{2}}=
	\frac{4\rho_{q}\rho_{s}k_{0\,z}\mathrm{Re}\tilde{k}_{q}}
	{\left|\rho_{q}k_{0\,z}
	+\rho_{s}\tilde{k}_{q}\right|^{2}}. 
\end{equation} 
The most interesting case for us is the interface between superfluid helium and a solid. In this case, as shown above, $N\!=\!3$. Due to smallness of the parameters $\rho_{q}/\rho_{s},s_{i}/s_{0}\!\ll\!1$ for $i\!=\!1,2,3$, the transmission coefficient is obtained in the effective limit $\tilde{k}_{s}\!\rightarrow\!0$, so $D\!\ll\!1$. For $S\!=\!2$ and $N\!=\!1$ the expression (\ref{D}) turns into the one obtained in \cite{JLTP2006}. For $S\!=\!N\!=\!3$ it turns into the result of \cite{PRB2008}.

Eq. (\ref{D}) is valid for any polynomial $\Omega^{2}(k^{2})$, monotonic or not, any incidence angle or any density ratio $\rho_{s}/\rho_{q}$. The structure of $D$ depends, through $\tilde{k}_{q}$, on the number of traveling waves $N$, which is in turn determined by (\ref{Snell}). For example, if, for some set of parameters, $\omega$ and incidence angle, Snell's law (\ref{Snell}) ensures that there are no traveling waves in the quantum fluid $N\!=\!0$, then $\tilde{k}_{q}$ is imaginary and $D\!=\!0$. If $N\!>\!1$, then the transmitted energy flow is divided between several traveling waves and we need to find the partial transmission coefficients which give the shares of the full energy of the incident wave, that is transferred to each of the newly created waves.

\begin{figure}[htb]
\center
\includegraphics[width=8.6cm, viewport=111 305 484 546]{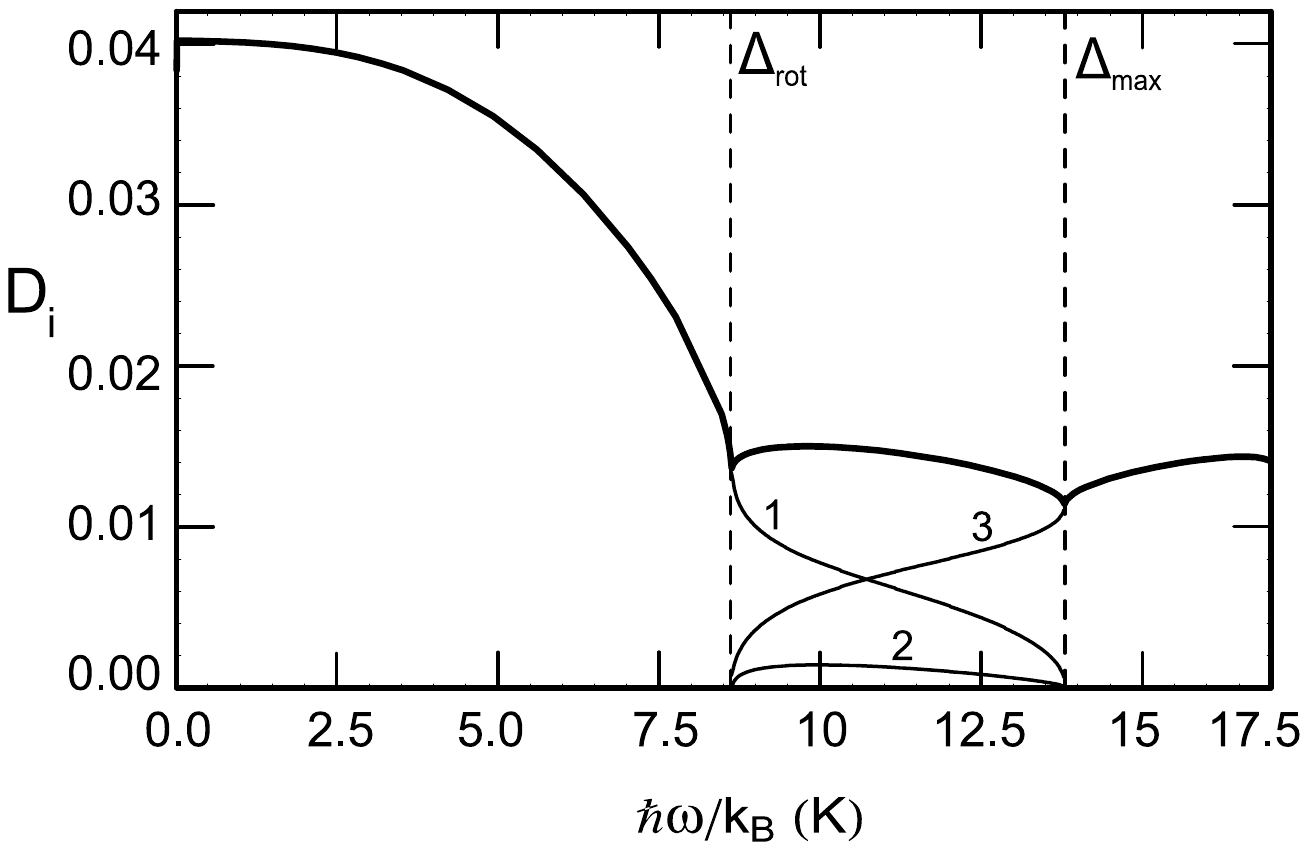}
\begin{minipage}[b]{0.85\textwidth}\caption{\label{Fig3}
Transmission coefficients for a phonon in the solid, at normal incidence to the interface, as functions of frequency $\omega$ (in Kelvin units), for typical values of parameters $\rho_{q}/\rho_{s}\!=\!s/s_{0}\!=\!0.1$. The full transmission coefficient $D$ is shown by the thick line, and the partial coefficients $D_{1,2,3}$ are marked by numbers $1$, $2$ and $3$ respectively.
These curves, as well as all the subsequent ones, are calculated for polynomial $\Omega^{2}(k^{2})$ of degree $S\!=\!21$, obtained by fitting the experimental data for $\Omega^{2}(k^{2})/k^{2}$ as function of $k^{2}$ by a polynomial of degree $S\!-\!1$.}
\end{minipage}
\end{figure}

Fig.\ref{Fig3} shows $D(\omega,0)$ for typical parameters of a helium-solid interface $\rho_{q}/\rho_{s},s/s_{0}\!=\!0.1$ for the full range of frequencies $0\!<\!\omega\!<\!17$K. We see, that at frequencies at which $N$ changes, the structure of $D$ changes and the curve has kinks. These and all the subsequent plots for the transmission and reflection coefficients are built for the polynomial $\Omega^{2}(k^{2})$ of degree $S\!=\!21$, obtained by fitting the experimental data \cite{neutron} for $\Omega^{2}(k^{2})/k^{2}$ as function of $k^{2}$ by a polynomial of degree $S\!-\!1$. The use of different approximation polynomials, which were obtained by varying the sampling of the experimental data, gives the curves that are indistinguishable on the given scale.

Fig.\ref{Fig4} shows the part of the previous graph in the range of energies $(\Delta_{rot},\Delta_{max})$. Comparing with the results of \cite{PRB2008}, we see that the qualitative behavior of the probability curves is the same, but the numerical values are smaller by about $\sim 20$\%.

\section{Energy flows and the partial transmission coefficients} 
As noted above, at the interface between a solid and superfluid helium the case $N\!=\!3$ is usually expected to be realized. This means that the out-solution contains three traveling waves, with $k_{1\,z}$, $k_{2\,z}$ and $k_{3\,z}$ (\ref{KzSeq}), which correspond to the phonon, $R^-$ roton and $R^+$ roton respectively. In order to find the partial transmission coefficients, corresponding to those waves, we need to calculate the energy flows in each of them.

Also the confirmation of energy conservation in the interaction process at the interface, is not only of abstract interest. Indeed, the first attempt to apply the nonlocal hydrodynamic description of a superfluid to the problem of quasiparticles interaction with the interface, was probably made in work \cite{P&S92}, in which the interaction of rotons with the surface excitations of He~II was studied. In that work the solution of the nonlocal wave equation in the half-space, analogous to Eq. (\ref{EQP}), was sought in the form of even functions of $z$, while extending the integration limits to infinity. There appeared no contradiction, as the reflection coefficient of a roton from the free surface had absolute value of $1$ either way. However, when applied by the authors to the problem of two adjacent media, such approach gave physically irrelevant solutions, in which energy was not conserved to the extent of the non-linearity of the dispersion relation of one of the media. The reason for this is that integral equations cannot be ordinarily solved by even (or odd) continuation. One can see from Eq. (\ref{EQP}), that even if the values of $P(z)$ at $z\!<\!0$ had physical sense, they would in fact be fully determined by its values at $z\!>\!0$, through the integral in the right-hand part. The demands for the equation to hold in $z\!<\!0$, and for the solution to be even (or odd), lead to non-physical nonlocal bonds between the points of the fluid in the physical region $z\!>\!0$ with the points in the non-physical region $z\!<\!0$.  This leads to a violation of the energy conservation law. This issue led the present authors to engage in solving Eq. (\ref{EQP}), with the integrand given on the half-line $z>0$ and with finite integration limits, using the  Wiener and Hopf method \cite{PRB}-\cite{PRB2008}. We solve it without assuming that the integrand holds at $z<0$.

Therefore let us calculate the energy flows in the solid and in the quantum fluid separately. First, the energy flow to the interface from the solid is the difference of the flows in the incident and the reflected waves. If we express the amplitudes through $P_{0}\equiv\left.P\right|_{z=-0}$, and use (\ref{D}), for the average $z$-component of the energy flow density we obtain
\begin{equation} \label{Flux-}
	 Q_{0\,z}=\frac{|P_{0}|^{2}}{2\rho_{0}\omega}\mathrm{Re}\,\tilde{k}_{q}.
\end{equation} 
Now we take into account that Eq. (\ref{DispEq}) is a polynomial equation with real coefficients with regard to $k_{z}^{2}$. Therefore its roots break up into real ones and into complex-conjugate pairs. The negative real roots give imaginary $k_{i\,z}$, and the complex-conjugate pairs $k_{i\,z}^{2}\!=\!(k_{j\,z}^{2})^{\ast}$ give the pairs of roots $k_{i\,z}$ in $\mathcal{C}_{+}$, that are anti complex-conjugate: $k_{i\,z}=-k_{j\,z}^{\ast}$. So all the non-real roots $k_{i\,z}$ do not give contributions to $\mathrm{Re}\tilde{k}_{q}$. The same consideration applies to the summands $k_{i\,z}(\omega\!=\!0)$, which are all complex. Therefore $\tilde{k}_{q}=\sum_{i=1}^{N}k_{i\,z}$ and both (\ref{D}) can be simplified and for $Q_{0\,z}$ we obtain 
\begin{equation} \label{Flux-f}
	 Q_{0\,z}=\frac{|P_{0}|^{2}}{2\rho_{0}\omega}\sum\limits_{i=1}^{N}k_{i\,z}.
\end{equation}

The expression is very simple and might seem trivial, but it turns out that thinking that the energy flow in the $i$-th wave in superfluid corresponds to the summand with subscript $i$ in (\ref{Flux-f}), and thus is proportional to $k_{i\,z}$, is completely wrong. Indeed, we may remember that $k_{2\,z}\!<\!0$, while each wave created in superfluid can only carry energy away from the interface, not towards it. Let us calculate the energy flows in each wave explicitly.

If $\alpha_{out,i}$ is the pressure amplitude of a traveling wave $i$, and its group velocity is $u_{i}$, then the average $z$-component of its energy flow density is 
\begin{equation} \label{Qi}
	 Q_{i\,z}=\frac{|k_{i}k_{i\,z}|}{2\rho_{0}\omega^{2}}
		\left|\alpha_{out,i}^{2}u_{i}\right|,\quad i=1,2,3. 
\end{equation}

The amplitudes $\alpha_{out,i}$ are calculated as the residues of $P_{out}$ (\ref{Solution}) in $k_{i\,z}$, and group velocities $u_{i}$ can be obtained from the following representation of the dispersion law 
\begin{equation}
\label{DispEqProduct}
	\Omega^{2}(k)=Ak^{2}\prod\limits_{i=2}^{S}
		\left(k_{z}^{2}-k_{i\,z}^{2}(\omega\!=\!0)\right),
\end{equation}
where $A$ is a constant and $k_{i\,z}$ the same roots as in (\ref{Knew}). After some transformations and using the boundary conditions (\ref{BC}) we can obtain
\begin{equation} \label{ur2}
	\alpha_{out,j}^{2}u_{j}=P_{0}^{2}\frac{\omega}{k_{j}}
	\prod\limits_
	{\genfrac{}{}{0pt}{}{i=1}{i\neq j}}^{S}
	\left\{\frac{k_{j\,z}+k_{i\,z}}{k_{j\,z}-k_{i\,z}}
	\cdot\frac{k_{j\,z}-k_{i\,z}(\omega=0)}
		{k_{j\,z}+k_{i\,z}(\omega=0)}\right\},\quad j=1,2,3. 
\end{equation}

Now we take into account the structure of the roots $k_{i\,z}$, i.e that they are either real, or imaginary, or break up into pairs related as $k_{i\,z}=-k_{j\,z}^{\ast}$. Therefore for the case $N=3$ we have
\begin{equation} \label{PartialFluxes}
	Q_{i\,z}=\frac{|P_{0}|^{2}}{2\rho_{0}\omega}\cdot
	 k_{i\,z}\frac{k_{i\,z}+k_{j\,z}}{k_{i\,z}-k_{j\,z}}
		\frac{k_{i\,z}+k_{k\,z}}{k_{i\,z}-k_{k\,z}}, 
		\quad\{i,j,k\}\!=\!\{1,2,3\}+perm., 
\end{equation} 
where $perm.$ denotes all permutations. We can see that for any $i\!=\!1,2,3$ the energy flow $Q_{i\,z}$ is positive, but not proportional to $k_{i\,z}$. It can be now shown easily from (\ref{PartialFluxes}), that in all the cases $N\!\leq\!3$ 
\begin{equation} \label{EnergyOK} 
	\sum\limits_{i=1}^{N}Q_{i\,z}=Q_{0\,z}. 
\end{equation} 
So energy is conserved and the solution is consistent.

The relative energy flows in the three waves are functions of only $k_{1\,z},k_{2\,z}$ and $k_{3\,z}$. This is very important. All the complex roots $k_{i\,z}$ are obtained by constructing an approximation polynomial $\Omega^{2}(k^{2})$ for the experimental data for the dispersion curve, then finding all of the roots of Eq. (\ref{DispEq}) on the complex plane, then sorting out those of them that lie in $\mathcal{C}_{+}$. Those are the necessary steps in order to, for example, build the graph $D(\omega,k_{\tau})$ given by Eq. (\ref{D}) in Fig.\ref{Fig3}, as the complex roots enter the expressions explicitly. The functions $k_{i\,z}(\omega,k_{\tau})$ differ slightly for different approximations, with the most rough values given by the case $S=3$ (see \cite{PRB2008}).

However, now, in order to calculate the relative energy flows in the phonon, $R^-$ roton and $R^+$ roton waves, based on Eq. (\ref{PartialFluxes}), all we need is the experimental data for $k_{1,2,3}(\omega)$, and the result is not affected by the approximation polynomial.

The partial transmission coefficients for $i=1,2,3$ are $D_{i}=DQ_{i\,z}/Q_{0\,z}$. In the most interesting case $N\!=\!3$ then we obtain 
\begin{equation} \label{Dpartial} 
	 D_{i}=D\cdot
	\frac{k_{i\,z}}{k_{i\,z}\!+\!k_{j\,z}\!+\!k_{k\,z}}
	\frac{k_{i\,z}\!+\!k_{j\,z}}{k_{i\,z}\!-\!k_{j\,z}}
		\frac{k_{i\,z}\!+\!k_{k\,z}}{k_{i\,z}\!-\!k_{k\,z}},
	\quad\{i,j,k\}\!=\!\{1,2,3\}+perm. 
\end{equation}

\begin{figure}[!ht]
\center
\includegraphics[width=8.6cm , viewport=113 310 500 546]{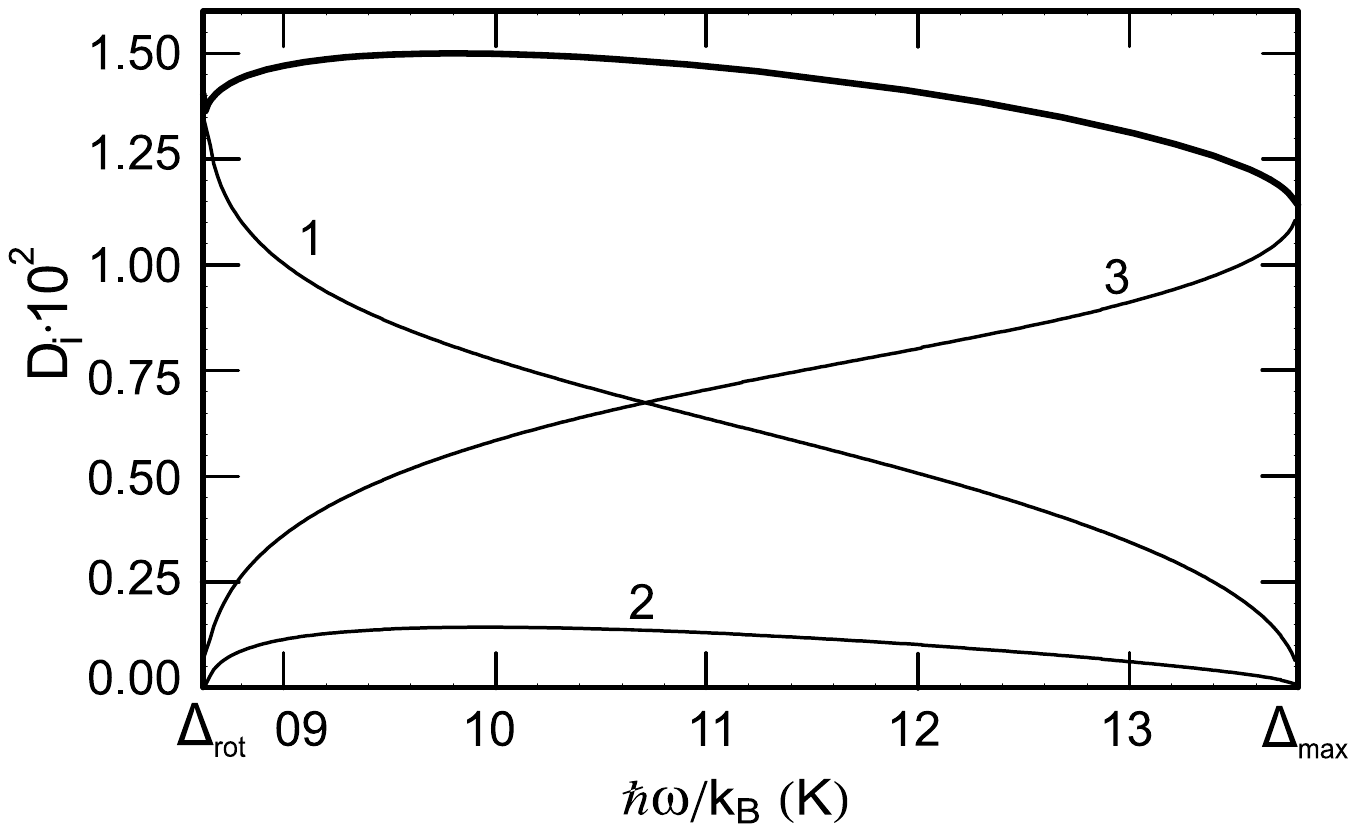}
\begin{minipage}[b]{0.85\textwidth}\caption{\label{Fig4}
An enlargement of the Fig.\ref{Fig3} in the range $\omega\!\in\!(\Delta_{rot},\Delta_{max})$, showing greater detail. Note that the ratios $D_{i}/D_{j}$ depend only on $k_{1,2,3}$.}
\end{minipage}
\end{figure}

The asymptotes of the frequency dependencies of the transmission coefficients,  at $\omega$ close to the roton minimum $\Delta_{rot}$ or maxon maximum $\Delta_{max}$, are determined by the factors that are functions of $k_{1,2,3\,z}$ only. When $\omega\!\rightarrow\!\Delta_{rot}\!+\!0$, the small parameter is $\tilde{\omega}\!=\!\omega\!-\!\Delta_{rot}$, and $(k_{2\,z}\!+\!k_{3\,z})\!\sim\sqrt{\tilde{\omega}}\!\rightarrow\!0$, therefore $D_{2,3}\!=\!O(\tilde{\omega})$. Likewise when $\omega\!\rightarrow\!\Delta_{max}\!-\!0$, we obtain $D_{1,2}\!=\!O(\sqrt{\Delta_{max}\!-\!\omega})$ (see Fig.\ref{Fig4}).

At normal incidence $k_{\tau}=0$ and $k_{1,3\,z}=k_{1,3}$, while for the $R^-$ rotons we have $k_{2\,z}=-k_{2}$. Therefore, from (\ref{Dpartial}) we obtain
\begin{eqnarray} \label{DpartialNorm} 
	&D_{1}(\omega,0)\sim k_{1}
		\frac{\displaystyle k_{2}\!-\!k_{1}}{\displaystyle k_{2}\!+\!k_{1}}
		\frac{\displaystyle k_{3}\!+\!k_{1}}{\displaystyle k_{3}\!-\!k_{1}};\quad
	 D_{3}(\omega,0)\sim k_{3}
		\frac{\displaystyle k_{3}\!+\!k_{1}}{\displaystyle k_{3}\!-\!k_{1}}
		\frac{\displaystyle k_{3}\!-\!k_{2}}{\displaystyle k_{3}\!+\!k_{2}};
	\nonumber \\
	&D_{2}(\omega,0)\sim k_{2}
		\frac{\displaystyle k_{2}\!-\!k_{1}}{\displaystyle k_{2}\!+\!k_{1}}
		\frac{\displaystyle k_{3}\!-\!k_{2}}{\displaystyle k_{3}\!+\!k_{2}}. 
\end{eqnarray}

In (\ref{DpartialNorm}) both fractions in $D_{2}$ are less than unity, due to the negative dispersion of the $R^-$ rotons, which implies $k_{2\,z}\!<\!0$, as opposed to the analogous expressions for $D_{1,3}$, in which one of the fractions is always grater than $1$. This is the reason that $D_{2}<D_{1,3}$. The graphs of $D_{i}/D$ for normal transmission are shown on Figs.\ref{Fig3} and \ref{Fig4}, and there it is seen clearly that the difference in partial transmission coefficients is significant: 
\begin{equation} \label{D2} 
	 D_{2}\ll D_{1,3}.
\end{equation}

The graphs on Figs. \ref{Fig3},\ref{Fig4} display the same qualitative behaviour of $D_{i}(\omega,0)$ as the analogous graphs obtained in \cite{PRB2008}, and the two sets look very similar to each other, which shows that the preliminary results of \cite{PRB2008} were correct.

Due to the strong inequality $s_{sol}\gg s_{i}$, and the generalized Snell law (\ref{Snell}), the wave vectors of the traveling waves of superfluid helium are concentrated in a narrow cone around the normal: $\sin\Theta_{i}\ll\sin\Theta_{sol}$, and so $\Theta_{i}\ll \pi/2$ for $i=1,2,3$. Therefore in the zero approximation of the small parameter $s/s_{sol}$, the angular dependencies of transmission coefficients are reduced to the multiplier $\cos\Theta_{0}\sim k_{0\,z}$ 
\begin{equation} \label{DpartialTheta}
	 D_{i}(\omega,\Theta_{0})\approx D_{i}(\omega,0)\cos\Theta_{0} 
\end{equation}
and thus Eq. (\ref{D2}) holds for any angles of incidence.

In terms of quasiparticles this result means that a phonon in the solid, incident on the interface with superfluid helium, creates $R^-$ rotons with much smaller probability than it creates phonons and $R^+$ rotons in the helium (those probabilities are small themselves, too, due to the small coefficient of transmission of the wave at the interface). Due to the law of detailed balance, likewise, when an $R^-$ roton is incident on the interface with a solid, a phonon in the solid is created with much less probability than when a phonon or $R^+$ roton in the helium, is incident.

Thus, in an experiment in which quasiparticles of superfluid helium are created by a solid heater and registered on a solid bolometer, $R^-$ rotons are weakly created and weakly detected (where "weakly"\, means much smaller than helium phonons or $R^+$ rotons), and the effect gets squared. This result explains why $R^-$ rotons were not detected in direct experiments until the work \cite{exp2}, in which both creation and detection of quasiparticles was carried out by different methods. There, the $R^-$ rotons were created by collisions in the bulk liquid. 

\section{He~II quasiparticles incident on the interface}
 Let us consider the problem of a helium quasiparticle of type $j$ incident on the interface. It can be a phonon $j=1$, $R^-$ roton $j=2$ or $R^+$ roton $j=3$.  First, we discuss the consequences of Snell's law (\ref{Snell}) with regard to this case, assuming, as before, the set of inequalities 
\begin{equation} \label{Sounds}
	 s_{0}>s_{1}>s_{2}>s_{3}>0. 
\end{equation}

If a wave $j$ is incident on the interface at angle $\Theta_{j}$ to the normal, the value of $k_{\tau}=k_{j}\sin\Theta_{j}$ is set and the angles of propagation of other waves $i\neq j$ are determined by Eq. (\ref{Snell}): $\sin\Theta_{i}=s_{i}/s_{j}\,\sin\Theta_{j}$. For $i\!<\!j$ we can define six critical angles 
\begin{equation} \label{Critical}
	\sin\Theta_{ji}^{cr}=\frac{s_{j}}{s_{i}}\quad j>i;\;i,j=0,1,2,3. 
\end{equation}
If the incidence angle of wave $j$ is greater than $\Theta_{ji}^{cr}$, it means that $k_{\tau}\!>\!k_{i}(\omega)$, so $k_{i\,z}$ is imaginary, the $i$-th wave is damped and the corresponding quasiparticle is not created.

The solution in $z\!>\!0$, which corresponds to this problem, should contain one traveling wave $j$, which travels towards the interface (or, rather, the wave packet constructed of such waves should be traveling towards the interface; this makes a difference for $R^-$ rotons). Such a solution of form (\ref{Solution}) is easily constructed based on the out-solution $P_{out}$.

Let us take the solution of the form (\ref{Solution}) and use the same rules (\ref{Kz-signs}) of roots shifting from the real line as used in construction of $P_{out}$, except for the pair $\pm k_{j\,z}$, which will be shifted in the opposite directions. Thus the wave packet $j$ traveling away from the interface is replaced by the wave packet of the same type traveling towards the interface, and the notation differs from $P_{out}$ by the sign of $k_{j\,z}$ (only in the denominator, as the roots $k_{i\,z}$ in the numerator of (\ref{Solution}) do not change sign). Let us denote this new solution as $P_{in}^{(j)}$. We can construct $N$ such solutions, after the number of real pairs of roots $\pm k_{i\,z}$ of Eq. (\ref{DispEq}): 
\begin{equation} \label{Pin} 
	 P_{in}^{(j)}=\left.P_{out}\right|_{k_{j\,z}
		\rightarrow -k_{j\,z}},\quad j=1,\ldots N. 
\end{equation} 
The amplitude of $P_{in}^{(j)}$ is $C_{in}^{(j)}$, and the amplitude of pressure in each monochromatic wave in it is $\alpha_{in,i}^{(j)}$, in full analogy to the notations for $P_{out}$.

Now, for the case $N\!=\!3$ we have four solutions of form (\ref{Solution}), $P_{out}$ and $P_{in}^{(j)}$ for $j=1,2,3$, which are linearly-independent due to their structure. The full number of solutions that can be formed by picking different triplets out of the real six roots to be shifted into $\mathcal{C}_{+}$ is $\binom{3}{6}=20$ (where $\binom{n}{k}$ is the binomial coefficient). In the case $S\!=\!N\!=\!3$ it was shown explicitly \cite{PRB2008}, that all of them can be presented as linear combinations of the four built above. The solutions of Eq. (\ref{EQP}) with the corresponding kernel were sought in the form (\ref{SolutionR}), and it was shown that the space of solutions is four-dimensional. In the general case the additional complex roots $k_{i\,z}$ with $i\!>\!3$ should not affect this circumstance, and we assume that, in quantum-mechanical terms, the four solutions form the basis set  for the level at  given $\omega$ and $k_{\tau}$. The degeneracy of the level is due to the non-monotonicity of the dispersion relation $\Omega(k)$. The actual dimension is $(N\!+\!1)$ and depends on $\omega$ and $k_{\tau}$.

The basis is constructed conveniently in such a way, that when a wave $i$ is incident, the solution in $z\!>\!0$ is $(P_{out}\!+\!P_{in}^{(i)})$.

The case $N=0$ does not correspond to the considered problem. It is realized when there are no real roots $k_{i\,z}$: $k_{\tau}\!>\!k_{1}(\omega)$ at $\omega\!<\!\Delta_{rot}$ or $k_{\tau}\!>\!k_{3}(\omega)$ at $\omega\!>\!\Delta_{max}$. Then the solution $P_{out}$ is exponentially damped with $z$. It either corresponds to the surface excitation of superfluid helium, ripplon, or to the complete internal reflection of a  phonon in the solid from the interface with He~II, which is usually not realized, as mentioned above, due the small sound velocity of superfluid helium. The are no in-solutions and the level with $\omega$ and $k_{\tau}$ is not degenerate.

The case $N=1$ is realized when there is only one pair of real $k_{i\,z}$: a) $k_{\tau}\!<\!k_{1}(\omega)$ at $\omega\!<\!\Delta_{rot}$, b) $k_{3}(\omega)\!>\!k_{\tau}\!>\!k_{2}(\omega)$ at $\Delta_{rot}\!<\omega\!<\!\Delta_{max}$, c) $k_{\tau}\!<\!k_{3}(\omega)$ at $\omega\!>\!\Delta_{max}$. The cases a) and c) correspond to a phonon or $R^+$ roton incident on the interface when there are no quasiparticles of other types with the same frequency. The case b) corresponds to an $R^+$ roton incident on the interface at angles greater than the second critical $\Theta_{32}^{cr}$, so the phonon and $R^-$ rotons waves have imaginary $k_{i\,z}$ and these quasiparticles cannot be created. This situation is qualitatively equivalent to the problem of the interaction of a quasiparticle with monotonic dispersion with the interface. The level is doubly degenerate, the same as in the usual case, when there are just the incident and the reflected waves. The basis is $\{P_{out},\,P_{in}^{(j)}\}$, with $j=1$ for a) and $j=3$ for b) and c).

The case $N=2$ is realized when there are two pairs of real $k_{i\,z}$: $k_{1}(\omega)\!<\!k_{\tau}\!<\!k_{2}(\omega)$ at $\Delta_{rot}\!<\omega\!<\!\Delta_{max}$. It corresponds to an $R^+$ or an $R^-$ roton incident on the interface at angles greater than $\Theta_{31}^{cr}$ and $\Theta_{21}$ respectively. Thus the phonon wave has imaginary $k_{i\,z}$ and is damped. The basis is $\{P_{out},\,P_{in}^{(2)},\,P_{in}^{(3)}\}$ and the solution is $P_{out}\!+\!P_{in}^{(j)}$ with $j=2$ when the $R^-$ roton is incident and $j=3$ when the $R^+$ roton.

Finally, the case $N=3$ is realized when the number of real roots $k_{i\,z}$ is a maximum: $k_{\tau}\!<\!k_{1}(\omega)$ at $\Delta_{rot}\!<\omega\!<\!\Delta_{max}$. This corresponds to a quasiparticle $i$ of He~II incident on the interface at angles smaller than $\Theta_{ij}^{cr}$ for all $j\!>\!i$. So all the three waves $i=1,2,3$ in the helium are traveling waves. The basis is $\{P_{out},\,P_{in}^{(1)}\,P_{in}^{(2)},\,P_{in}^{(3)}\}$ and the solution is $P_{out}\!+\!P_{in}^{(j)}$ with $j=1,2$ or $3$ depending on the type of the incident wave.

So, when a wave $i$ is incident on the interface, the solution in the quantum fluid is $P_{out}\!+\!P_{in}^{(i)}$. Thus it consists of one incident wave $i$, $N$ reflected waves and $S-N$ damped waves, with two free amplitudes $C_{out}$ and $C_{in}^{(i)}$ (the last $S$ waves are present in both $P_{out}$ and $\!P_{in}^{(i)}$, so their amplitudes are summed together). The ratio $C_{in}^{(i)}/C_{out}$ is obtained with the help of the boundary conditions. The values of pressure $P_{in}^{(i)}$ and the normal component of velocity $\mathrm{v}_{in\,z}^{(i)}$ in the $P_{in}^{(j)}$-solution are calculated in the same way as for $P_{out}$, and we obtain 
\begin{equation} \label{PVin0}
	\left.\frac{\mathrm{v}_{z}}{P}\right|_{z=+0}=
	\frac{1}{\rho_{0}\omega}\frac{C_{out}\tilde{k}_{q}
		+C_{in}^{(i)}\tilde{k}_{q}^{(i)}}{C_{out}+C_{in}^{(i)}},
\end{equation}
where
\begin{equation} \label{kQi}
	\tilde{k}_{q}^{(i)}=\left.\tilde{k}_{q}\right|_{k_{i\,z}
	\rightarrow -k_{i\,z}}=\tilde{k}_{q}-2k_{i\,z}. 
\end{equation}

If $\Theta_{i}\!<\!\Theta_{i0}^{cr}$, then the solution in the solid, in $z\!<\!0$, is one transmitted traveling wave with the normal component of wave vector $k_{0\,z}'\!=\!-k_{0}\cos\Theta_{0}\!<\!0$, where $\Theta_{0}$ is determined from (\ref{Snell}), so we have
\begin{equation} \label{PVin0-}
	\left.\frac{\mathrm{v}_{z}}{P}\right|_{z=-0}=\frac{1}{\rho_{s}\omega}
	k_{0\,z}'= -\frac{1}{\rho_{q}\omega}\tilde{k}_{s}.
 \end{equation} 
If $\Theta_{i}\!>\!\Theta_{i0}^{cr}$ then the wave in the solid is damped and $k_{0\,z}'=-i|\sqrt{k_{\tau}^{2}\!-\!k_{0}^{2}}|$. In this case we define $\tilde{k}_{s}$ such that  Eq. (\ref{PVin0-}) is valid again: i.e. $\tilde{k}_{s}\!=\!-\frac{\rho_{q}}{\rho_{s}}k_{0\,z}'$. Due to the strong inequality $s_{0}\!\gg\!s_{i}$, $\Theta_{i0}^{cr}\!\ll\!1$ and the first variant is realized only for helium quasiparticles incident in a narrow cone around the normal to the interface.

Applying the boundary conditions (\ref{BC}) to (\ref{PVin0}) and (\ref{PVin0-}), we obtain the ratio $C_{in}^{(i)}/C_{out}$ and thus express the amplitudes of all the waves through the amplitude of the incident one. The amplitude reflection coefficient $r_{ij}$ is the ratio of pressures in the created wave $j$ and the incident wave $i$: 
\begin{eqnarray} \label{rii}
	 r_{ii}&=&-
	\prod\limits_{j=2}^{S}\frac{k_{i\,z}\!-\!k_{i\,z}(\omega\!=\!0)}
	{k_{i\,z}\!+\!k_{i\,z}(\omega\!=\!0)}\cdot
	\prod\limits_
	{\genfrac{}{}{0pt}{}{j=1}{j\neq i}}^{S}
	\frac{k_{i\,z}\!+\!k_{j\,z}}{k_{i\,z}\!-\!k_{j\,z}}\cdot
	\frac{\tilde{k}_{s}\!+\!\tilde{k}_{q}\!-\!2k_{i\,z}}
	{\tilde{k}_{s}+\tilde{k}_{q}};\\
\label{rij}
	 r_{ij}&=&-\frac{2k_{i\,z}}{k_{j\,z}-k_{i\,z}}\cdot
	\frac{\tilde{k}_{s}\!+\!\tilde{k}_{q}\!-\!(k_{i\,z}\!+\!k_{j\,z})}
	{\tilde{k}_{s}+\tilde{k}_{q}},\qquad i\neq j. 
\end{eqnarray}

The energy reflection coefficient $R_{ij}$ is the ratio of the normal components of the energy fluxes in the created wave packet $j$ and the incident wave packet $i$. It is also the creation probability of quasiparticle $j$ when quasiparticle $i$ is incident on the interface. In other words $R_{ij}$ is the probability of mode change from $i$ to $j$, so it can be also called the conversion coefficient. For $i\!=\!j$ we have $R_{ii}\!=\!|r_{ii}|^{2}$, and for $i\!\neq\!j$ we use amplitude coefficients $r_{ij}$ and group velocities (\ref{ur2}). Most of the factors have absolute value $1$ in the similar way to the transition from Eq. (\ref{Qi}) to Eq. (\ref{PartialFluxes}), and after some simplifications we obtain 
\begin{eqnarray} \label{Rii}
	 R_{ii}&=&
	\left|\frac{(k_{i\,z}\!+\!k_{j\,z})(k_{i\,z}\!+\!k_{k\,z})}
		{(k_{i\,z}\!-\!k_{j\,z})(k_{i\,z}\!-\!k_{k\,z})}\right|^{2}\cdot
	\left|\frac{\tilde{k}_{s}+\tilde{k}_{q}-2k_{i\,z}}
		{\tilde{k}_{s}+\tilde{k}_{q}}\right|^{2}; \\
\label{Rij}
	 R_{ij}&=&
	\frac{|4k_{i\,z}k_{j\,z}|}{(k_{i\,z}\!-\!k_{j\,z})^{2}}\,
	\left|\frac{(k_{i\,z}\!+\!k_{k\,z})(k_{j\,z}\!+\!k_{k\,z})}
		{(k_{i\,z}\!-\!k_{k\,z})(k_{j\,z}\!-\!k_{k\,z})}\right|\cdot
	\left|\frac{\tilde{k}_{s}\!+\!\tilde{k}_{q}\!-\!(k_{i\,z}\!+\!k_{j\,z})}
		{\tilde{k}_{s}+\tilde{k}_{q}}\right|^{2},\\ &&\mbox{where}\quad
	\{i,j,k\}=\{1,2,3\}+perm.\nonumber 
\end{eqnarray}
We see that $R_{ij}\!=\!R_{ji}$.

The formulae (\ref{Rii}) and (\ref{Rij}) are universal, in the sense that they are applicable to all possible cases $N=1,2,3$ and at any angles of incidence, provided $k_{i\,z}$ and $k_{j\,z}$ are real (otherwise it either does not have sense or $R_{ij}\!=\!0$). For example, let us find the reflection coefficient for the $R^+$ roton incident at angle $\Theta_{3}\!>\!\Theta_{32}^{cr}$. In this case $\{i,j,k\}=\{3,1,2\}$, $k_{3\,z}$ is real but $k_{0\,z}$, $k_{1\,z}$ and $k_{2\,z}$ are imaginary. So only $R_{33}$ has sense. The first term in (\ref{Rii}) is equal to unity. The real parts of the numerator and denominator of the second fraction are equal to $-k_{3\,z}$ and $k_{3\,z}$ correspondingly, while the rest of the summands are imaginary (see transition from Eq. (\ref{Flux-}) to Eq. (\ref{Flux-f})) and give equal imaginary parts of the numerator and denominator. So the second term is also equal to unity and we obtain the obvious result $R_{33}\!=\!1$. Likewise, the expression for $R_{ij}$ (\ref{Rij}) is simplified when $k_{k\,z}$ is imaginary (the case $N\!=\!2$) -- the middle term equals $1$.

If $\Theta_{i}\!<\!\Theta_{i0}^{cr}$, then $\tilde{k}_{s}$ is real and the transmission coefficient is $D_{i}=1-\sum\limits_{j=1}^{3}R_{ij}$ (due to the law of detailed balance it is the same $D_{i}$, as functions of the conserved quantities $\omega$ and $k_{\tau}$, in Eq. (\ref{Dpartial})), otherwise $\tilde{k}_{s}$ is complex and $D_{i}\!=\!0$. The energy conservation law has the form 
\begin{equation} \label{Rij-D}
	\sum\limits_{j=1}^{3}R_{ij}+D_{i}=1\quad\mbox{for}\quad \forall i\leq N,
\end{equation} 
including the cases when some of the coefficients are equal to zero. It was verified in \cite{PRB2008} explicitly in all cases for $S\!=\!3$.

At the interface between superfluid helium and a solid $\tilde{k}_{s}\!\ll\!k_{i}$ for all $i$, so all the transmission coefficients (\ref{D}) and (\ref{Dpartial}) are small $D,D_{i}\!\ll\!1$ and the reflection coefficients $r_{ij}$ (\ref{rii},\ref{rij}) and $R_{ij}$ (\ref{Rii},\ref{Rij}) are obtained effectively in the limit $\tilde{k}_{s}\!\rightarrow\!0$.

The frequency dependencies of $R_{ij}$ and their asymptotes close to $\Delta_{rot}$ and $\Delta_{max}$ are determined by the factors in (\ref{Rii},\ref{Rij}), that are functions only of $k_{1,2,3\,z}$. At $\tilde{\omega}\!=\!\omega\!-\!\Delta_{rot}\!\rightarrow\!+0$ we have $(k_{2\,z}\!+\!k_{3\,z})\!=\!O(\tilde{\omega}^{1/2})$, and therefore from (\ref{Rii}) and (\ref{Rij}) we obtain 
\begin{eqnarray} \label{RijAsympt1}
	&R_{11}=1-D_{1}-O(\tilde{\omega}^{1/2});\quad
	 R_{23}=1-O(\tilde{\omega}^{1/2});&\\ \label{RijAsympt2}
	&R_{22,33}=O(\tilde{\omega});\qquad\qquad\quad
	 R_{12,13}=O(\tilde{\omega}^{1/2}).& 
\end{eqnarray} 
So the dominating processes on the interface near the roton minimum are reflection of phonons into phonons, conversion of $R^-$ rotons into $R^+$ rotons and visa versa.

\begin{figure}[!htb]
\center
\includegraphics[width=8.6cm, viewport=89 297 510 559]{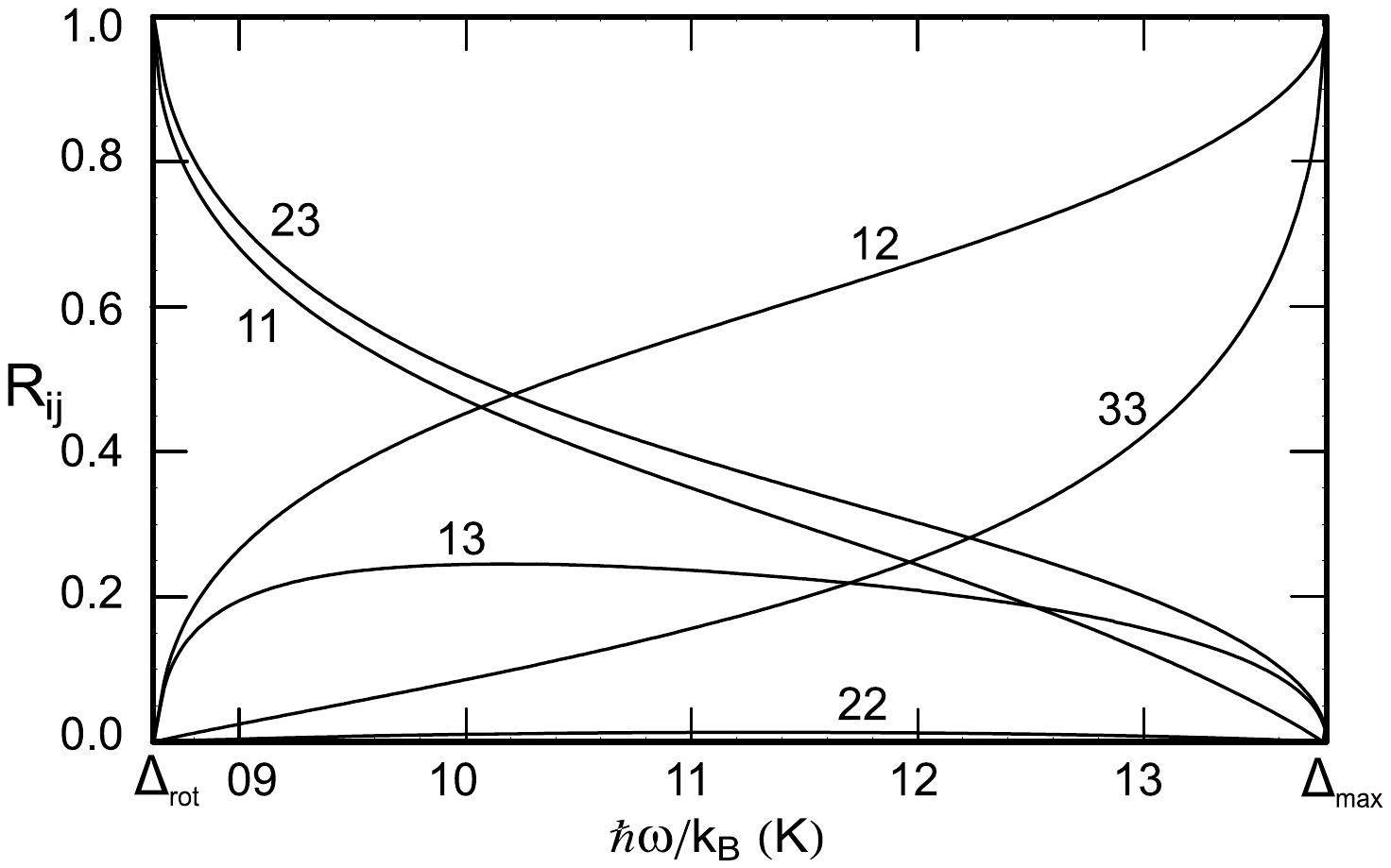}
\begin{minipage}[b]{0.85\textwidth}\caption{\label{Fig5}
Reflection and conversion coefficients $R_{ij}$ at normal incidence as functions of $\omega$, for typical parameters $\rho_{q}/\rho_{s}\!=\!s/s_{0}\!=\!0.1$. The curves are denoted by their
respective pairs of subscripts $ij$.}
\end{minipage}
\end{figure}

Likewise we can derive the asymptotes of $R_{ij}$ at $\omega\!\rightarrow\!\Delta_{max}\!-\!0$ for $k_{\tau}\!<\!k_{1,2}$. They can be obtained from (\ref{RijAsympt1}) and (\ref{RijAsympt2}) by changing the subscripts $\{1,2,3\}$ to $\{3,1,2\}$ and the small parameter $\tilde{\omega}$ to $\Delta_{max}\!-\!\omega$. The graphs of $R_{ij}$ as functions of frequency for $k_{\tau}\!=\!0$, are shown on Fig.\ref{Fig5}.

\begin{figure}[!ht]
\center
\includegraphics[width=8.6cm, viewport=118 315 471 537]{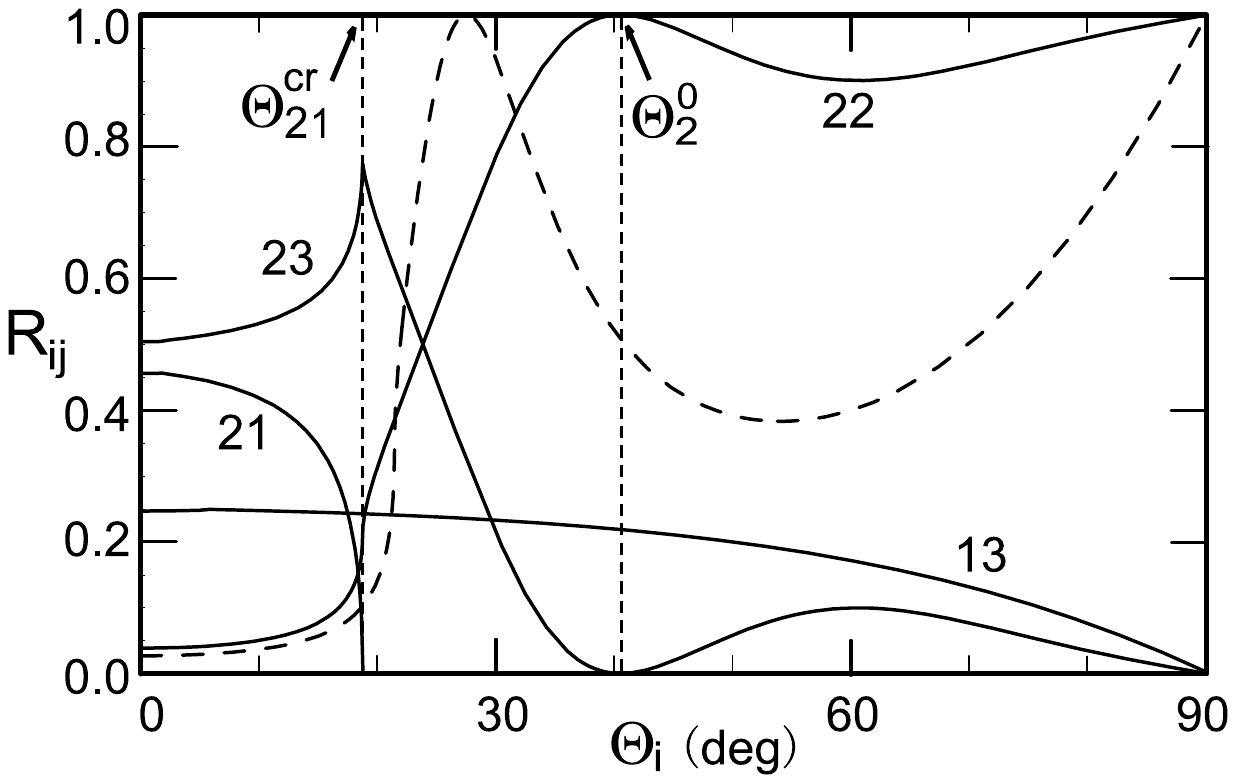}
\begin{minipage}[b]{0.85\textwidth}\caption{\label{Fig6}
Coefficients $R_{2j}$ and $R_{13}$ as functions incidence angles $\Theta_{2}$ and $\Theta_{1}$ respectively, $\hbar\omega\!=\!k_{B}\cdot10$ K, $\rho_{q}/\rho_{s}\!=\!s/s_{0}\!=\!0.1$. The
curves are denoted by their respective pairs of subscripts $ij$, and the dashed line shows $R_{22}$ obtained in \cite{PRB2008}.}
\end{minipage}
\end{figure}

The angular dependencies of $R_{ij}(\omega,\Theta_{i})\!=\!R_{ij}(\omega,k_{\tau}\!=\!k_{i}\sin\Theta_{i})$ are nontrivial even in the case of the same small parameters $\rho_{q}/\rho_{s}$ and $s/s_{0}$, because the angles $\Theta_{i}$ now are not necessarily small. All the reflection coefficients are related through the energy conservation law (\ref{Rij-D}), and therefore can be expressed through the coefficients $D_{1,2,3}$ and $R_{21,22,23,31}$. The graphs of the four reflection coefficients as functions of the corresponding angles of incidence,  at a fixed frequency $\hbar\omega\!=\!k_{B}\cdot10$K and $\rho_{q}/\rho_{s}\!=\!s/s_{0}\!=\!0.1$, are shown on Fig.\ref{Fig6} and illustrate the peculiarities of angular dependencies of $R_{ij}$.

Comparing with the previous results \cite{PRB2008}, we see that the qualitative behavior of the probability curves is the same, but the numerical values differ by $\sim10$\%, with the exception of the curves for $R_{11}$ and $R_{33}$. The probability $R_{11}$ is on average almost twice as large as its preliminary value, and $R_{33}$ is almost twice as small. This difference is clearly caused by the roughness of the approximation for the $R^+$ roton branch in \cite{PRB2008}.

From (\ref{Rij}) we see that $R_{ij}\!\sim\!k_{i\,z}k_{j\,z}$ for $i\!\neq\!j$. Therefore for $i\!>\!j$, when the incidence angle is close to $\pi/2$, $k_{i\,z}\!\rightarrow\!0$ and the coefficient $R_{ij}$ tends to zero as $\sqrt{\pi/2\!-\!\Theta_{i}}$; for $i\!<\!j$, when the incidence angle is close to the critical angle $\Theta_{ij}^{cr}$, $k_{j\,z}\!\rightarrow\!0$ and the coefficient $R_{ij}$ tends to zero as $\sqrt{\Theta_{ij}^{cr}\!-\!\Theta_{i}}$. This gives the asymptotic behaviour of $R_{12,13,23}$ at  angles of incidence close to $\pi/2$ and of $R_{21,31,32}$ at incident angles close to the corresponding critical angles ($R_{ij}$ and $R_{ji}$ are the same functions of $\omega$ and $k_{\tau}$ but different functions of respective incident angles $\Theta_{i}$ and $\Theta_{j}$). We see that at grazing incidence, the only coefficients that do not tend to zero are $R_{ii}$ and so the dominating processes are specular (mirror) reflections; in the limit $\tilde{k}_{s}\!\rightarrow\!0$ the transmission coefficients are zero and so $R_{ii}\!\rightarrow\!1$.

When the incident angle for some quasiparticle $i$ increases and becomes greater that one of the critical angles $\Theta_{ij}^{cr}$ ($j\!<\!i$), the root $k_{j\,z}$ changes from being real to imaginary. Therefore the structure of the coefficients $R_{ij}$ changes even for those of them that do not tend to zero or unity. Thus, if we consider an $R^-$ roton incident, when $\Theta_{2}\!=\!\Theta_{21}^{cr}$, $R_{21}$ tends to zero and $R_{22,23}$ have kinks.

The numerator of a factor of $R_{23}$ in (\ref{Rij}) is $k_{23}=\tilde{k}_{q}\!-\!k_{2\,z}-\!k_{3\,z}$, equal to $k_{1\,z}-k_{2\,z}(\omega\!=\!0)-k_{3\,z}(\omega\!=\!0)+\sum_{i=4}^{S}(k_{i\,z}-k_{i\,z}(\omega\!=\!0))$. As $k_{1\,z}$ changes from being real to imaginary, the real part of $k_{23}$ turns to zero. In the most simple case $S\!=\!3$, which was analyzed in \cite{PRB2008}, the imaginary part without $k_{1\,z}$ is $2\mathrm{Im}k_{2\,z}(\omega\!=\!0)$ and is of the order of the exponent of the one-dimensional kernel $h(z;\mathbf{k}_{\tau})$, which replaces $h(\mathbf{r})$ in Eq. (\ref{EQP}) after Fourier transforms by $x$ and $y$. This exponent determines the effective correlation length in the $z$-direction for given $k_{\tau}$. It was shown, that there is an angle $\Theta_{2}\!=\!\Theta_{2}^{0}$ a little greater than $\Theta_{21}^{cr}$, for which $k_{23}$ tends to zero, and this corresponds to the penetration depth of the phonon wave $|k_{i\,z}|^{-1}$ being of the order of the effective one-dimensional correlation length in the quantum fluid. For this angle $\Theta_{2}^{0}$ we have $R_{21}\!=\!D_{2}\!=\!0$ because $\Theta_{2}^{0}\!>\!\Theta_{21}^{cr}\!>\!\Theta_{20}^{cr}$, and also $R_{23}\!=\!0$. Therefore $R_{22}\!=\!1$ and the $R^-$ rotons are specularly reflected. The existence of such an angle enabled the authors of Ref. \cite{CondMat} to suggest an experiment for detection of $R^-$ rotons by measuring the negative pressure that an $R^-$ roton beam would exert on a membrane. The absolute value of the pressure is greatest when the $R^-$ rotons are specularly reflected from it, without mode change, and this condition is satisfied for the incidence angle equal to $\Theta_{2}^{0}$.

In the general case, the effective one-dimensional kernel is a sum of many exponents, which are now also  functions of $\omega$, but the combination of $k_{i\,z}$ that enters the numerator $k_{23}$ still plays the role of some effective exponent. So $R_{22}$ will turn to unity if the penetration depth of the phonon wave becomes equal to the corresponding effective correlation length. We can see in Fig.\ref{Fig6}, that this angle $\Theta_{2}^{0}$ indeed exists for the $\Omega(k)$ that well approximates the dispersion relation of superfluid helium, as opposed to the rough approximation with $S\!=\!3$ used in \cite{CondMat}. Its value is about $40^{\circ}$, as opposed to the previous result $\sim\!30^{\circ}$, the peak is not as sharp, and at larger angles $R_{22}$ is quite close to unity. This means that we only need to provide $\Theta_{2}\!>\!40^{\circ}$ for the negative pressure of the $R^-$ roton beam to be measurable, though best results would be at $\Theta_{2}\!\approx\!40^{\circ}$.

Comparing the graphs in Fig.\ref{Fig6} with the analogous graphs of \cite{PRB2008} shows, that at angles less than the critical angle the difference is small, but when the incidence angle is greater than the critical angle, the difference increases. For comparison, the dashed line shows $R_{22}$ from \cite{PRB2008}.

If we consider reflection from the free surface, then the boundary conditions (\ref{BC}) are replaced by (\ref{boundary-landau}). For a solution $\sim\exp(i\mathbf{k}_{\tau}\mathbf{r}_{\tau}\!-\!i\omega t)$, this expression turns into $P\!=\!-\sigma k_{\tau}^{2}\xi$. The $z$-component of velocity of the surface in this case is $\mathrm{v}_{z}\!=\!-i\omega\xi$, and therefore we have 
\begin{equation} \label{BCfree}
	 \left.\mathrm{v}_{z}\right|_{z=0}=
	\frac{i\omega}{\sigma k_{\tau}^{2}} \left.P\right|_{z=0}. 
\end{equation} 
Thus the ratio $\mathrm{v}_{z}/P$ at $z\!=\!0$ is given not by (\ref{PVin0-}), but by (\ref{BCfree}). Therefore the reflection coefficients from the free surface of superfluid helium are obtained by formally replacing $\tilde{k}_{s}$ in Eqs. (\ref{rii}-\ref{Rij}) by 
\begin{equation} \label{Kf}
	\tilde{k}_{f}=-i\frac{\rho_{q}}{\sigma}\frac{\omega^{2}}{k_{\tau}^{2}}.
\end{equation} 
The quantity $\tilde{k}_{f}$ is imaginary, so the structure of the coefficients is the same as when $D\!=\!0$, and $\sum_{j=1}^{3}R_{ij}=1$.

The transition layer of helium atoms at the interface can be treated in the usual approach as multiple layers of continuous media with corresponding equilibrium densities and equations of state. This turns the problem into one of passing waves through this multilayered structure.

Then the solution in the outmost ``layer'', the bulk fluid, is composed of $P_{out}$, $P_{in}^{(1)}$, $P_{in}^{(2)}$ and $P_{in}^{(3)}$. The one thing that changes is the boundary conditions on the interface between this ``layer'' and the next, which alters the relative amplitudes of the four solutions in the particular solution. This is due to changes in $Z$ (\ref{Z}), or equivalently, $\tilde{k}_{s}$ (\ref{Knew}) (or $\tilde{k}_f$ in (\ref{Kf})), so the transmission (\ref{D}) and reflection (\ref{rii}-\ref{Rij}) coefficients also will change through the replacement of $Z$ by some effective quantity.

However, when considering the process of a solid's phonon incident on the interface, the solution in the fluid can only be $P_{out}$, as it is the unique eigensolution that describes the fluid's reaction to outside influence (with new effective $Z$). So the relative transmission coefficients, i.e. the ratios $D_{2}/D_{1}$ and $D_{2}/D_{3}$, which are functions only of $k_{1,2,3}$, stay exactly the same as derived in this paper and Eq. (\ref{Dpartial}) holds exactly quantitatively. Thus the result on the weak creation and detection of $R^{-}$ rotons also holds.

Likewise there are no changes to the derived asymptotes and qualitative behavior of the reflection coefficients $R_{ij}$, as well as the critical angles $\Theta_{ij}^{cr}$, which depend only on the frequency and angular dependences of $k_{1,2,3\,z}$, and thus only on the dispersion relation of the bulk fluid.

The influence of the microstructures of the interface can be taken into account by introducing the effective roughness of a solid surface. Estimates were made in \cite{AdamenkoSurf1,AdamenkoSurf2} (for phonons) and show that this roughness does not change the results significantly.

It can be noted, that it is also possible to take into account the layer of weakly-adsorbed atoms on the interface, see \cite{Syrkin}. Incorporation of this analysis into our model would also lead to a change of the effective impedance $Z$ (\ref{Z}), in the way similar to the multilayered problem discussed above, with analogous consequences.

\section{Conclusion} 
In the current work we solve the problem of the interaction of superfluid helium phonons and rotons with interfaces in a consistent and unified way. We describe the quantum fluid in the dispersive (nonlocal) hydrodynamics approach, in which a medium's dispersion relation $\Omega(k)$ enters the equations explicitly and serves as the only input "parameter". We present the consistent solution of the problem of the interaction of quasiparticles with the interface, for the case when their dispersion relation is arbitrary and nonmonotonic, so that $\Omega^{2}(k^{2})$ is a polynomial of some degree $S$.

We solve the equations of nonlocal hydrodynamics in the half-space and discuss some general consequences of the solution's structure with regard to the problem of quasiparticles creation on interface. These are a generalization of Snell's law (\ref{Snell}) with phase velocities as functions of frequency, the realization of backward reflection and backward refraction for $R^-$ rotons, existence of multiple critical angles corresponding to each pair of modes in the two the adjacent media.

The creation probabilities, at the interface, of each quasiparticle of both media, phonons in the solid and phonons, $R^-$ rotons and $R^+$ rotons in the superfluid helium,  are derived when any of them are incident on the interface (\ref{D}), (\ref{Dpartial}), (\ref{Rii}), (\ref{Rij}). The obtained expressions are valid for all frequencies, below and above the roton gap, any incident angle, and effectively arbitrary dispersion relation of the quantum fluid. The peculiarities and asymptotic behaviour of the probabilities as functions of frequency and angles are analysed.

This work includes and generalizes the results of Refs. \cite{JLTPold}-\cite{PRB2008}, in which different specific special cases of dispersion relation were considered, and in the corresponding special cases, the expressions obtained here turn into the ones obtained earlier, so they are now presented from a single point of view. All the qualitative results of \cite{PRB2008} with regard to superfluid helium are confirmed, including the explanation why $R^-$ rotons were not detected in experiments until \cite{exp2}. For the suggestion of new experiments on $R^-$ rotons detection, we have adjusted the optimal angle of incidence. The general expressions obtained for the transmission and reflection coefficients allowed us to refine the dependencies obtained in \cite{PRB2008} for the superfluid helium - solid interface.

We hope that the obtained results stimulate new experiments on the interaction of helium phonons and rotons with interfaces.

\section*{Appendix. Wiener and Hopf method} 
Let us show how the integro-differential equation (\ref{EQP}) is brought to a Riemann boundary value problem (also called Riemann-Hilbert problem). The idea of this transition is called the Wiener and Hopf method.

We start from Eq. (\ref{EQP}) and make Fourier transform by time:
\begin{equation} \label{EQPomega} 
	\triangle P(\mathbf{r},t)=
	-\omega^{2} \!
	\int\limits_{z_{1}>0}\! d^{3}r_1\,h(|\mathbf{r}\!-\!\mathbf{r}_1|)P(\mathbf{r}_1,t),
	\quad x,y,t\!\in\!(0,\infty),\;z\!\in\!(0,\infty). 
\end{equation} 
Let us introduce two new functions $P^{+}(\mathbf{r},t)$ and $P^{-}(\mathbf{r},t)$. The first one is defined as 
\begin{eqnarray} \label{A-P->}
	&P^{-}(\mathbf{r},t)=P(\mathbf{r},t)&\mbox{for}\quad z>0\\
\label{A-P-<}
	\mbox{and}&P^{-}(\mathbf{r},t)=0&\mbox{for}\quad z<0. 
\end{eqnarray}

Likewise we demand for the second function 
\begin{equation} \label{A-P+>}
	 P^{+}(\mathbf{r},t)=0\qquad\mbox{for}\quad z>0, 
\end{equation} 
and at $z<0$ it will be defined so that the equation 
\begin{equation} \label{A-P+<}
	\triangle(P^{-}(\mathbf{r}) +P^{+}(\mathbf{r}))=
	-\omega^{2}\int\limits_{V}\,d^{3}r_{1} h(|\mathbf{r}_{1}-\mathbf{r}|)P^{-}(\mathbf{r}_{1}), 
\end{equation}
where integration now is made over the infinite space, should hold true on $z<0$. At $z>0$ it holds automatically.

Now the right-hand part of Eq.  (\ref{A-P+<}) is a convolution, so after Fourier transform by $\mathbf{r}$ 
\begin{equation} \label{A-PpmK}
	 P^{\pm }(\mathbf{k})=\int\, d^{3}r\, e^{-i\mathbf{kr}} P^{\pm } (\mathbf{r}).
\end{equation} 
we have 
\begin{equation} \label{A-Riemann}
	 P^{-}(\mathbf{k})
	\frac{\Omega^{2}(k)-\omega^{2}}
		{\Omega^{2}(\mathbf{k})}+P^{+}(\mathbf{k})=0. 
\end{equation}

We are interested in $P^{\pm}$ as functions of $k_{z}$, while $\omega$ and $\mathbf{k}_{\tau}\!=\!\mathbf{e}_{x}k_{x}\!+\!\mathbf{e}_{y}k_{y}$ act as parameters. Due to the demands (\ref{A-P->}) and (\ref{A-P+>}), for the functions $P^{\pm}(z)$ that grow not faster than power law at infinity, their Fourier images (\ref{A-PpmK}) $P^{+}(k_{z})$ and $P^{-}(k_{z})$ are analytical in the upper and lower half-plane of the complex plane of variable $k_{z}$ respectively. So the upper index indicates the half-plane of the complex plane $k_{z}$, in which the respective function is analytical. Eq. (\ref{A-Riemann})
holds for real $k_{z}$, so in detailed notation we have 
\begin{equation}\label{A-RiemannF}
	 P^{-}(k_{z};\omega,\mathbf{k}_{\tau})
	\frac{\Omega^{2}(k_{z},\mathbf{k}_{\tau})-\omega^{2}}
	{\Omega^{2}(k_{z},\mathbf{k}_{\tau})}+P^{+}(k_{z};\omega,\mathbf{k}_{\tau})=0,
		\quad k_{z}\in(-\infty,\infty). 
\end{equation}

This equality gives a linear relation on the real line between the limit values of functions $P^{+}(k_{z})$  and $P^{-}(k_{z})$ on it, which are analytical in the upper and lower half-planes of the complex variable $k_{z}$ respectively. Therefore it defines a homogeneous Riemann boundary value problem on the real line (see for example \cite{Gahov}) with "density" 
\begin{equation}
\label{A-densityG}
	 G(k_{z};\omega,k_{\tau})=\frac{\Omega^{2}(k)-\omega^{2}}{\Omega^{2}(k)}.
\end{equation} 
Its key parameter is the density's index, which can be calculated as difference between the number of plain zeros of Eq. (\ref{DispEq}) in the upper half-plane $\mathcal{C}_{+}$ of $k_{z}$ and the lower half-plane $\mathcal{C}_{-}$. As the function $\Omega^{2}(k_{z})$ is even, the index of $G$ (\ref{A-densityG}) is equal to zero.

The Riemann boundary problem with density $G$, that is differentiable on the contour (in our case it is the real line), does not tend to zero on it and has zero index, has a unique solution to within a multiplicative constant, provided that there is an additional condition on its asymptotic behaviour at infinity \cite{Gahov}. Our problem has two complications with regard to this standard case. First, $G$ has zeros in the real roots of Eq. (\ref{DispEq}), and second, it is unbounded at zero when $k_{\tau}\!=\!0$.

The detailed solution of this problem in the case of monotonic dispersion is given in appendix to \cite{PhNT}, in the one-dimensional case, when $k_{\tau}\!=\!0$. It was shown there, that the singularity of $G$ in zero changes the structure of the solution, which is still unique if we demand that $P(z)$ is bounded at  $z\!>\!0$. As the three-dimensional solution has to turn into the one-dimensional one when $k_{\tau}\!=\!0$, it has the same structure, notwithstanding the fact that the singularity takes place for a single value of the parameter $k_{\tau}$.

The simplest way to bypass the first complication is to shift the real roots into the complex plane, while preserving the index of density $G$. When $\Omega(k)$ is monotonic, there are two real roots and so two ways of shifting while preserving the index. Thus we obtain two linear-independent solutions, that play the same role as plane waves with positive and negative $k_{z}$ in the usual case of linear dispersion. When $\Omega(k)$ in non-monotonic, the situation is more complex. In \cite{PRB2008} the Eq. (\ref{EQP}), with the corresponding kernel, was solved explicitly by using a different method, for the special case when $\Omega^{2}(k^{2})$ is a cubic polynomial and there are three positive roots of Eq. (\ref{DispEq}) with regard to $k_{z}^{2}$. It was shown there, that there are four linear-independent solutions with given $\omega$ and $\mathbf{k}_{\tau}$, and the general solution is their linear combination. It can be also shown, that these four solutions can be obtained by the Wiener and Hopf method by selecting appropriately which three of the six real roots of Eq. (\ref{DispEq}) with regard to $k_{z}$ should be shifted up (or, equivalently, down) from the real line. All the rest of the $\binom{3}{6}\!=\!20$ solutions obtained this way are linear combinations of the four. As the structure and dimension of solutions of Eq. (\ref{EQP}) with given $\omega$ and $\mathbf{k}_{\tau}$ can only depend on the number of possible traveling waves in them, we use the same scheme of roots shifting to derive all the linear-independent solutions of (\ref{EQP}) in the case of arbitrary $\Omega(k)$, which leads us to (\ref{Solution}).

\acknowledgements{We are grateful to EPSRC of the UK (grant EP/F 019157/1) for supporting this work.}

\end{document}